\documentclass[
reprint,
superscriptaddress,
nofootinbib,
amsmath,
amssymb,
aps,
prd,
]{revtex4-2}

\usepackage[english]{babel}

\usepackage{graphicx}
\usepackage{amsfonts}
\usepackage{mathrsfs}
\usepackage{bm}
\usepackage{mathtools}
\usepackage{nccmath}
\usepackage{bigints}

\usepackage{enumerate}
\usepackage{multirow}
\usepackage{booktabs}
\usepackage{comment}
\usepackage{subfigure}
\usepackage{placeins}

\usepackage{soul}
\usepackage[normalem]{ulem}
\usepackage{physics}
\usepackage{setspace}

\usepackage[usenames,dvipsnames]{xcolor}

\usepackage[
colorlinks,
urlcolor=blue,
citecolor=blue,
linkcolor=blue
]{hyperref}

\allowdisplaybreaks

\begin{document}
\title{A parametrized test of general relativity for inspiralling eccentric binaries in LISA }
\author{Pankaj Saini}
\email{pankaj.saini@nbi.ku.dk}
\affiliation{Niels Bohr International Academy, The Niels Bohr Institute, Blegdamsvej 17, DK-2100, Copenhagen, Denmark}
\affiliation{Center of Gravity, Niels Bohr Institute, Blegdamsvej 17, 2100 Copenhagen, Denmark}
\author{Sylvain Marsat}
\email{sylvain.marsat@l2it.in2p3.fr}
\affiliation{Univ Toulouse, CNRS, L2IT, Toulouse, France}
\author{Lorenz Zwick}
\affiliation{Niels Bohr International Academy, The Niels Bohr Institute, Blegdamsvej 17, DK-2100, Copenhagen, Denmark}
\affiliation{Center of Gravity, Niels Bohr Institute, Blegdamsvej 17, 2100 Copenhagen, Denmark}
\author{J\'{a}nos Tak\'{a}tsy}
\affiliation{Institut für Physik und Astronomie, Universität Potsdam, Haus 28, Karl-Liebknecht-Str. 24-25, Potsdam, Germany}
\author{Johan Samsing}
\affiliation{Niels Bohr International Academy, The Niels Bohr Institute, Blegdamsvej 17, DK-2100, Copenhagen, Denmark}
\affiliation{Center of Gravity, Niels Bohr Institute, Blegdamsvej 17, 2100 Copenhagen, Denmark}

\begin{abstract}
The upcoming space-based detector Laser Interferometer Space Antenna (LISA) will observe inspiralling black hole binaries in the mHz band, many of which may retain measurable orbital eccentricity. In contrast to quasicircular binaries, eccentric systems radiate through multiple orbital harmonics, while relativistic periastron precession introduces an additional secular phase structure in the waveform. We exploit this structure to develop a parametrized test of general relativity (GR) for eccentric bound orbits. We construct a parametrized frequency-domain eccentric waveform model in which a phenomenological deviation parameter $\delta\alpha$ modifies the GR prediction for the conservative azimuthal-to-radial frequency ratio, with $\delta\alpha=0$ corresponding to GR. We consider two distinct parametrizations of this deformation. In the first parametrization, $\delta\alpha$ modifies only the explicit precession-dependent sideband structures of the waveform. In the second parametrization, the secular precession phase is assigned to the dominant lower-order angular carriers and retained in resummed form. The latter produces a substantially stronger response because the dominant waveform components coherently accumulate the modified phase. We implement both parametrized models within \textsc{lisabeta} and
perform a Bayesian analysis that includes the time and
frequency-dependent LISA response and the associated
time-delay-interferometry observables. We find that LISA can place stringent constraints on the parametrized deviation. For a binary with chirp mass $3000 M_{\odot}$,
initial eccentricity $e_0=0.5$ observed for four years at an SNR of
$50$, the second model yields a $90\%$ credible bound
of $|\delta\alpha|\lesssim \, 10^{-4}$. Increasing the eccentricity further sharpens the constraints by introducing additional harmonic structure and reducing degeneracies among the binary parameters. The framework developed here is general and can be extended to more complete eccentric waveform models.
\end{abstract}

\maketitle

\section{Introduction}
The Laser Interferometer Space Antenna (LISA)~\cite{2017arXiv170200786A} will open the mHz  gravitational-wave (GW) window and observe compact binaries across a broad range of masses and astrophysical environments, from massive black-hole binaries~\cite{PhysRevD.93.024003,Bellovary:2018gbb} and extreme-mass-ratio inspirals (EMRIs)~\cite{PhysRevD.73.064030, Moore:2017lxy} to the early inspiral of stellar-origin compact binaries~\cite{Liu:2014qaa, Lau:2019wzw, Andrews:2019plw,Chen:2020wan,Wagg:2021cst, Amaro-Seoane:2012vvq,Sesana:2016ljz}. A fraction of these binaries is expected to retain measurable orbital
eccentricity in the LISA band.

Eccentric compact binaries are especially interesting because eccentricity is a direct tracer of formation history. Isolated binaries are generally expected to have negligible eccentricity, whereas binaries formed in globular clusters, nuclear star clusters, AGN disks, or few-body captures can retain measurable eccentricity in ground-based detectors~\cite{Samsing:2013kua,PhysRevD.97.103014,Rasskazov:2019gjw,Zevin:2021rtf,DallAmico:2023neb, Rowan:2025xxb}. Several LIGO--Virgo--KAGRA events have also been reported as candidate eccentric binary-black-hole mergers, although the strength of the evidence remains event and waveform-model dependent~\cite{Romero-Shaw:2020thy,
Gayathri:2020coq,Romero-Shaw:2021ual, Romero-Shaw:2022xko,Gupte:2024jfe, Morras:2025xfu,Jan:2025fps,
Kacanja:2025kpr}. Dynamical processes can produce a significant population of eccentric stellar-mass binary black holes in the mHz GW band~\cite{Fang:2019dnh, Hoang:2019kye, Stephan:2019fhf, Wang:2020jsx,Xuan:2023azh, Romero-Shaw:2024klf}. Eccentric binaries provide a qualitatively richer harmonic content that can be used to test GR.

Gravitational-wave observations have enabled precision tests of GR in the strong-field regime. Most current null tests introduce phenomenological deformations of the quasicircular inspiral, merger, or
ringdown waveform and infer whether the corresponding deviation
parameters are consistent with zero~\cite{Will:2005va,Sathyaprakash:2009xs,Yunes:2013dva,Berti:2015itd,Krishnendu:2021fga}. To date the theory has passed all such tests with no evidence for deviation~\cite{LIGOScientific:2016lio,LIGOScientific:2018dkp,LIGOScientific:2020tif,LIGOScientific:2021sio,DuttaRoy:2024qbl, Mahapatra:2025cwk, LIGOScientific:2026qni,LIGOScientific:2026fcf,LIGOScientific:2026wpt}. The growing catalog of detections nonetheless continues to sharpen these constraints~\cite{LIGOScientific:2026oim}. 

One way to test gravity using GWs is the parametrized inspiral test of GR~\cite{Yunes:2009ke, Agathos:2013upa}. In this approach, the inspiral part of the frequency-domain GW phase is parametrized in terms of phenomenological deviation parameters. These deviation parameters are introduced as free parameters and capture a particular class of deviations from GR~\cite{Yunes:2016jcc, Seymour:2026bjg}. In GR, the deviation parameters vanish. LISA can constrain these deviations with exquisite precision.~\cite{PhysRevLett.116.241104,Gair:2012nm, Datta:2023muk, LISACosmologyWorkingGroup:2026zah, Piarulli:2025rvr}. 

Extending this framework to eccentric binaries is important both because unmodelled eccentricity can
bias standard GR tests~\cite{PhysRevD.106.084031, PhysRevD.107.024009, PhysRevD.109.084056, Gupta:2024gun} as well as binary parameters~\cite{PhysRevLett.112.101101, PhysRevD.105.023003, Divyajyoti:2023rht, PhysRevD.110.024002}. Moreover, eccentric GW inspirals probe a different part of the conservative Hamiltonian of the system. The impact of eccentricity on tests of gravity depends on the source and the form of the non-GR correction. At small eccentricity,
additional parameter covariances can weaken the measurement, whereas at moderate eccentricity the richer harmonic content can break these degeneracies and improve the resulting bounds
~\cite{Ma:2019rei, Moore:2020rva}. Consistently modeling eccentricity has also been shown to strengthen theory-specific constraints and prevent false apparent deviations from GR in the analysis of GW200105~\cite{Roy:2025xih}.

In an eccentric binary, the radial and azimuthal motions provide two distinct orbital phase scales. General relativity predicts a secular advance of the periastron, quantified by the conservative azimuthal-to-radial frequency ratio $K\equiv \Omega_\theta/\Omega_r =1+k_{\rm GR}.$ In post-Newtonian (PN) framework, at leading 1PN order for a binary with mass $M$ and eccentricity $e$, $k_{\rm GR} = 3\left(M\Omega_r\right)^{2/3}/(1-e^2)$,  so that the accumulated periastron-precession phase satisfies
$\dot{\gamma}_{\rm GR}=k_{\rm GR}\Omega_r$. We introduce the
parametrized deformation
\begin{equation}
    K_\alpha= 1+\left(1+\delta\alpha\right)k_{\rm GR},
\end{equation}
or, equivalently,
$\dot{\gamma}_\alpha=(1+\delta\alpha)\dot{\gamma}_{\rm GR}$, with $\delta\alpha=0$ corresponding to GR. By construction, $\delta\alpha$ modifies only this conservative frequency relation, while the dissipative inspiral is held fixed to its GR prediction,
providing a controlled null test of the conservative dynamics. Although the geometrical location of periastron is not independently
defined for an exactly circular orbit, the frequency ratio $K$ has a circular limit that is distinct from one.
A deformation of this relation therefore produces a nonvanishing 1PN
phase correction even at zero eccentricity. Eccentricity makes the radial and azimuthal phase scales separately manifest through the orbital harmonics and
precession-dependent structures of the waveform, thereby improving the
measurability of $\delta\alpha$. 

Previous tests of GR using eccentric binaries have primarily used small-eccentricity
phasing models within PN framework. In particular, Ref.~\cite{PhysRevD.110.124062} introduced parametrized eccentric phasing coefficients and a deviation of the periastron-advance rate using
TaylorF2Ecc waveform model~\cite{PhysRevD.93.124061}. Reference~\cite{Bhat:2025lri} proposed a consistency test based on the evolution of orbital eccentricity to distinguish between eccentricity and other astrophysical, non-GR effects. Here we extend the parametrized test of GR beyond small-eccentricity phase corrections. \footnote{Recent work ~\cite{Chiaramello:2025bhi} has incorporated eccentricity and spin precession into inspiral--merger--ringdown parametrized effective-one-body tests,
including deviations in the late-inspiral dynamics and quasinormal-mode spectrum.}

For a controlled setup, we use a finite-eccentricity frequency-domain waveform complete through 1PN order in amplitude and phase and therefore restrict attention to non-spinning systems whose LISA-band evolution remains sufficiently far from merger for a weak-field description to be reliable. This setup allows us to isolate the conservative periastron-precession deformation without introducing additional higher-PN, spin, or merger--ringdown effects. The same construction can subsequently be incorporated into more complete eccentric waveform models and ultimately into multiband LISA–ground-based analyses, where the long LISA inspiral and the strong-field ground-based signal provide complementary information~\cite{Vitale:2016rfr, Chamberlain:2017fjl, Gnocchi:2019jzp, Carson:2019rda, Carson:2019kkh, Toubiana:2020vtf, Gupta:2020lxa,  Datta:2020vcj}. 

We perform Bayesian inference with the time- and frequency-dependent LISA response~\cite{PhysRevD.103.083011}, and explicitly examine how the inferred constraint depends on the placement of the deformation. We demonstrate that the placement of a phenomenological deformation within a finitely truncated waveform can change the apparent sensitivity by orders of magnitude. The last point is a general warning for theory-agnostic tests: a deviation
parameter is not fully specified until its projection onto the waveform basis is defined.

\begin{figure*}
    \centering
    \includegraphics[width=0.92\textwidth]{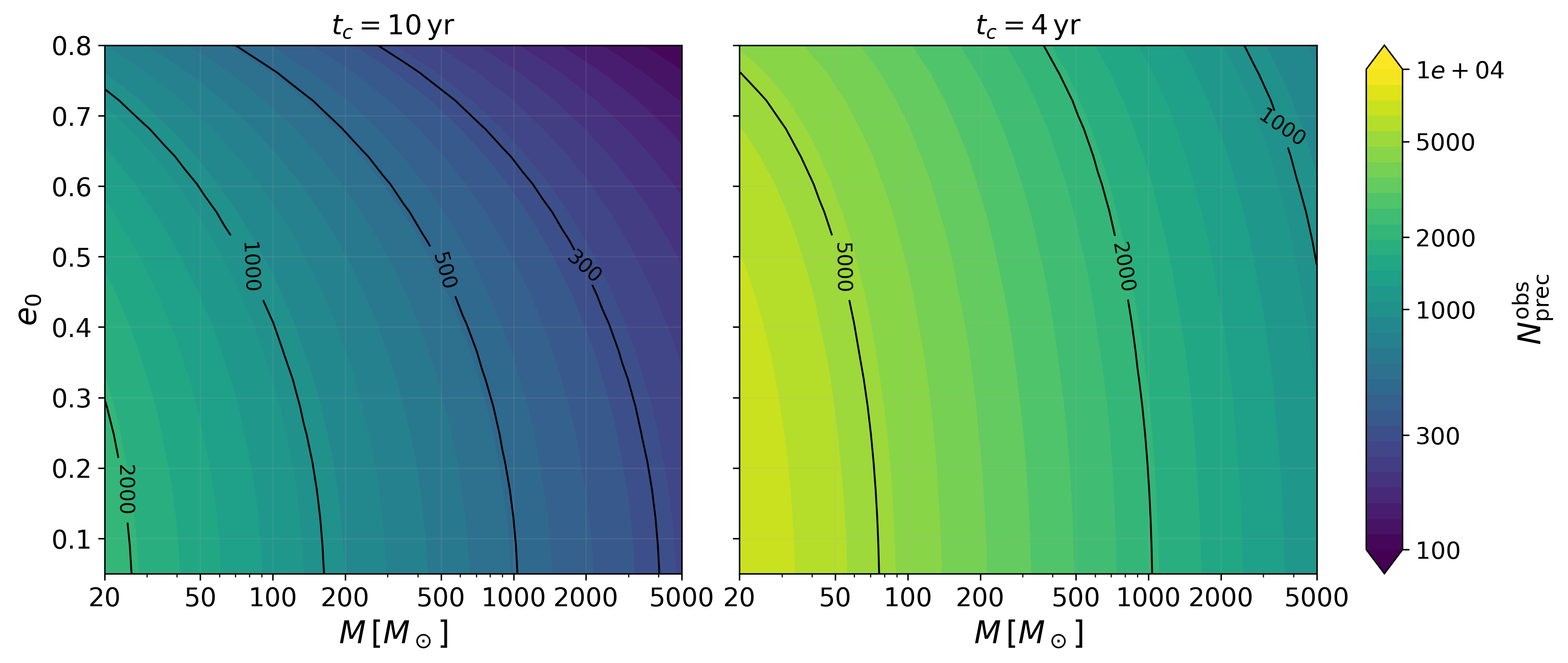}
    \caption{Number of periastron-precession cycles accumulated over the observed
portion of the signal, $N_{\rm prec}^{\rm obs}$, as a function of the total mass $M$ and initial eccentricity $e_0$ for equal-mass binaries. The two panels correspond to systems initialized at $t_c=10\,\mathrm{yr}$ (left) and $t_c=4\,\mathrm{yr}$ (right) before coalescence, with a fixed observation duration $T_{\rm obs}=4\,\mathrm{yr}$. The evolution is terminated at the smaller of the ISCO frequency and the adopted upper LISA frequency cutoff. Lower-mass binaries generally accumulate more precession cycles because they undergo more radial cycles during the observed interval. At fixed $M$ and $t_c$, increasing $e_0$ shifts the starting
frequency to lower values because eccentric binaries inspiral more rapidly. The resulting reduction in frequency competes with the explicit enhancement of the precession rate by $(1-e^2)^{-1}$, producing the eccentricity dependence seen in the figure.}
    \label{fig:precession}
\end{figure*}

\subsection{Accumulated periastron precession in the LISA band}
A useful diagnostic for identifying binaries with substantial periastron advance is the number of precession cycles accumulated over the observed portion of the signal,
\begin{equation}
    N_{\rm prec}^{\rm obs} =  \frac{1}{2\pi}
    \int_{t_{\rm in}}^{t_{\rm out}}
    \dot{\gamma}_{\rm GR}(t)\,dt .
    \label{eq:Nprec_obs}
\end{equation}
Here $t_{\rm in}$ and $t_{\rm out}$ determine the portion of the inspiral included in the observation.

Figure~\ref{fig:precession} shows $N_{\rm prec}^{\rm obs}$ for equal-mass binaries as a function of the total mass $M$ and initial eccentricity $e_0$. At each point in the $(M,e_0)$ plane, the starting
frequency is determined self-consistently from the 1PN eccentric evolution for a specified time to coalescence $t_c$. The left panel corresponds to binaries initialized $10\,\mathrm{yr}$ before
coalescence, whereas the right panel corresponds to binaries initialized $4\,\mathrm{yr}$ before coalescence. In both cases, the adopted observation duration is $T_{\rm obs}=4\,\mathrm{yr}$. For systems that reach the end of the inspiral within the observation window, the evolution is terminated at the smaller of the quadrupole
ISCO frequency and the adopted upper LISA cutoff.

At leading 1PN order, the GR periastron-precession rate is
\begin{equation}
    \dot{\gamma}_{\rm GR} =  6\pi\nu  \frac{(2\pi M\nu)^{2/3}}{1-e^2},
    \label{eq:gammadot_gr}
\end{equation}
where $\nu=\Omega_r/(2\pi)$ is the radial orbital frequency.
The quantity $\dot{\gamma}_{\rm GR}$ is the instantaneous precession rate, while the precession accumulated per radial orbit is
\begin{equation}
    \Delta\gamma_{\rm orb} = \frac{\dot{\gamma}_{\rm GR}}{\nu}   =
    2\pi k_{\rm GR}.
\end{equation}
At fixed radial frequency $\nu$, the precession rate $\dot{\gamma}_{\rm GR}$ increases with
eccentricity through the factor $(1-e^2)^{-1}$, while its frequency dependence scales as $\propto\nu^{5/3}$. However, the systems shown in Fig.~\ref{fig:precession} are not initialized at a common frequency. Instead, the starting frequency is chosen so that each binary has the specified time to coalescence. Because an eccentric binary evolves more rapidly than a circular one at the same frequency, increasing $e_0$ at fixed $M$ and $t_c$ generally shifts the starting frequency to smaller values. The resulting reduction in $\nu$ can compensate for, and in some regions outweigh, the explicit enhancement of the precession rate by
$(1-e^2)^{-1}$.

This effect is particularly apparent for the $t_c=10\,{\rm yr}$ case, where only the first $4\,{\rm yr}$ of the evolution are observed: systems with larger $e_0$ spend the observed interval at lower
frequencies and therefore need not accumulate more precession cycles despite their larger precession rate at fixed $\nu$. For $t_c=4\,{\rm yr}$, the observation extends to the end of the inspiral, so the binaries also sample the high-frequency part of the evolution where the precession rate grows rapidly. Consequently, the dependence of $N_{\rm prec}^{\rm obs}$ on $e_0$ is weaker than would be inferred from the factor $(1-e^2)^{-1}$ alone.

The quantity $N_{\rm prec}^{\rm obs}$ should therefore be interpreted as a diagnostic of the intrinsic precession-phase accumulation rather than as a complete forecast of parameter-estimation performance. Larger values of $N_{\rm prec}^{\rm obs}$ generally increase the distinguishability
of precession effects and can improve the sensitivity to
$\delta\alpha$. Since the dissipative evolution is held fixed, the
direct change in the accumulated precession phase is
\begin{equation}
    \gamma_\alpha-\gamma_{\rm GR} = 2\pi\delta\alpha\,N_{\rm prec}^{\rm obs}.
\end{equation}
The final constraint on $\delta\alpha$ also depends on the signal-to-noise ratio, the eccentric harmonic content, the time-dependent LISA response, and correlations with parameters such as
$M$, $e_0$, and the initial periastron phase.

The remainder of this paper is organized as follows.
In Sec.~\ref{sec:waveform-lisa-response}, we describe the basic
structure of the time-domain 1PN waveform and the orbital-evolution
equations within the generalized quasi-Keplerian parametrization. We then introduce the parametrized non-GR deformation of the conservative
frequency ratio, define the restricted- and full-model prescriptions, and construct the corresponding
frequency-domain waveforms using the stationary-phase approximation.
This section also presents the LISA response and the Bayesian-inference
setup implemented in \textsc{lisabeta}. In Sec.~\ref{sec:results}, we examine the distribution of signal power among the eccentric harmonics and precession sidebands and present the resulting constraints on $\delta\alpha$, including their dependence on the eccentricity and signal-to-noise ratio. We summarize our main conclusions in Sec.~\ref{sec:conclusion}. Appendix~\ref{app:hansen-mapping} describes the relation to the
generalized-Hansen representation. Appendix~\ref{app:harmonic-convergence} compares prescriptions for the maximum harmonic index required to capture $99\%$ of the signal power. Appendix~\ref{app:priors} lists the priors used in the analysis and presents the full-dimensional posterior distributions. Unless stated otherwise, we use units in which $G=c=1$.

\section{Waveform model and LISA response}
\label{sec:waveform-lisa-response}
We construct a frequency-domain eccentric inspiral waveform for LISA by combining 1PN eccentric dynamics~\cite{1PN_waveform}, a sideband stationary-phase approximation, and the full time-dependent LISA response. The purpose of the model is to define a controlled null test of the conservative periastron-precession sector. We keep the dissipative evolution of the orbital frequency and eccentricity fixed to its 1PN GR evolution, and introduce a single phenomenological parameter, $\delta\alpha$, that deforms the GR periastron-precession rate. The GR limit is recovered for $\delta\alpha=0$.

\subsection{1PN eccentric dynamics}
\label{subsec:onepn-dynamics}

\begin{figure}
    \centering
    \includegraphics[width=0.45\textwidth]{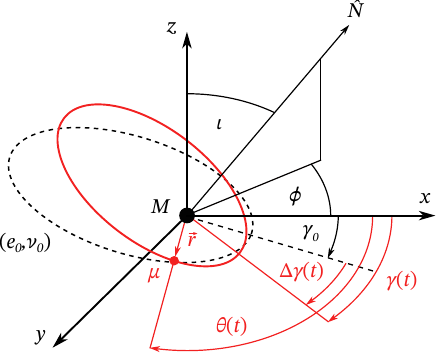}
    \caption{Schematic illustration of the eccentric binary geometry. The source-frame axes are denoted by $(x,y,z)$, with the line-of-sight direction $\hat{N}$ specified by the inclination $\iota$ and observer azimuth $\phi$. The total mass $M$ is shown at the origin and the reduced mass $\mu$ follows an eccentric orbit with separation vector $\Vec{r}$. The orbital phase angle is denoted by $\theta(t)$, while $\gamma(t)$ specifies the orientation of the pericenter in the orbital plane. The initial pericenter angle is $\gamma_0$, and relativistic pericenter precession accumulates an additional phase $\Delta\gamma$, so that $\gamma(t)=\gamma_0+\Delta\gamma (t)$.}
    \label{fig:schematic}
\end{figure}

Figure~\ref{fig:schematic} summarizes the source-frame geometry used in our eccentric waveform model. We adopt a Cartesian frame $(x,y,z)$ in which the orbital plane coincides with the $x$--$y$ plane and the $z$-axis is normal to the plane. The binary relative motion is described by a reduced mass $\mu$ moving on an eccentric orbit around the total mass $M$. The line of sight is described by the unit vector $\hat{N}$, parameterized by the inclination angle $\iota$ and the azimuthal angle $\phi$. In this convention,
\begin{equation}
\hat{N} = (\sin\iota \cos\phi, \sin\iota \sin\phi, \cos\iota).
\label{eq:line-of-sight}
\end{equation}
Here $\phi$ denotes the azimuth of the observer in the source frame and should not be confused with the generalized true anomaly $\varphi$ introduced below. The initial eccentric orbit is specified by the eccentricity $e_0$ and the initial radial orbital frequency $\nu_0$, or equivalently by the initial dominant quadrupole GW frequency $f_0= f_{22}^{\rm start} = 2\nu_0$. The orientation of the initial periastron is described by the periastron angle $\gamma_0$, measured from the positive $x$-axis. As the orbit precesses, the periastron advances by an angle $\Delta\gamma$, so that the instantaneous periastron direction is $\gamma (t)=\gamma_0+\Delta\gamma(t)$. This notation will be used throughout when discussing the orbital dynamics and the parametrized periastron-precession of the waveform.

We describe the eccentric orbit using the 1PN generalized
quasi-Keplerian parametrization~\cite{AIHPA_1985__43_1_107_0,Klein:2010ti}.
The radial separation is first written in terms of the eccentric anomaly
$u$ as
\begin{equation}
    r=a_r\left(1-e_r\cos u\right).
    \label{eq:radial-u}
\end{equation}
Introducing the generalized true anomaly $\varphi$, related to $u$ by
\begin{equation}
    \tan\frac{\varphi}{2}=\sqrt{\frac{1+e_r}{1-e_r}} \tan\frac{u}{2},
    \label{eq:true-anomaly}
\end{equation}
the radial motion and Kepler equation can equivalently be written as
\begin{equation}
    r= \frac{a_r\left(1-e_r^2\right)} {1+e_r\cos\varphi},
    \qquad
    l=u-e_t\sin u .
    \label{eq:radial-motion}
\end{equation}
Here $a_r$ is the radial semi-major axis, $e_r$ is the radial eccentricity, $e_t$ is the time eccentricity and $l$ is the mean anomaly. The radial orbital frequency is
\begin{equation}
\dot{l}=\Omega_r=2\pi\nu .
\label{eq:radial-frequency}
\end{equation}

At 1PN order the azimuthal motion contains a secular precession term and a periodic correction~\cite{1PN_waveform}, 
\begin{equation}
\theta-\theta_{0} = \left(1+\kappa_{1}\right)\varphi+\kappa_{2}\sin\varphi ,
\label{eq:theta-motion}
\end{equation}
with
\begin{equation}
\kappa_{1}=\frac{3M}{a_{r}(1-e_{r}^{2})}, \qquad \kappa_{2}=\frac{\mu e_{r}}{2a_{r}(1-e_{r}^{2})}.
\label{eq:kappa-definitions}
\end{equation}
The coefficient $\kappa_{1}$ gives the leading secular advance of the periastron, while $\kappa_{2}$ is a periodic 1PN correction to the angular motion. The distinction is important: after one radial period, $\varphi$ increases by $2\pi$, while $\sin\varphi$ returns to its initial value. Therefore, the $\kappa_{2}$ term does not contribute to the net secular advance over a radial cycle, whereas the $\kappa_{1}$ term gives an azimuthal advance $2\pi(1+\kappa_{1})$ per radial period. In what follows we use the radial eccentricity as the eccentricity variable and write $e\equiv e_r$. The time eccentricity $e_t$, which enters the Kepler equation, is not an independent parameter. It is fixed by the 1PN quasi-Keplerian relations~\cite{1PN_waveform}. Expanding the time eccentricity in terms of the radial eccentricity up to 1PN order gives
\begin{equation}
    e_t =e_r \left[
    1+ \left(4-\frac{3}{2}\eta\right)(2\pi M\nu)^{2/3}
    \right]+\mathcal O[(2\pi M\nu)^{4/3}],
    \label{eq:et-er-1pn}
\end{equation}
Thus $e_t=e_r$ in the Newtonian limit, while their difference is a
1PN correction controlled by the instantaneous orbital frequency $\nu$ and the symmetric mass ratio $\eta$. Along the inspiral, this relation is evaluated at the instantaneous pair $(e_r,\nu)$, so that the waveform is parameterized by a single eccentricity variable.

Expressing the secular precession parameter in terms of the radial frequency gives the GR azimuthal-radial frequency ratio
\begin{equation}
K_{\rm GR} \equiv \frac{\Omega_{\theta}}{\Omega_{r}} = 1+k_{\rm GR}.
\label{eq:k-definition}
\end{equation}
At leading 1PN order,
\begin{equation}
k_{\rm GR}(e,\nu) = \frac{3 \left(M\Omega_{r}\right)^{2/3}}{1-e^{2}},
\label{eq:kgr}
\end{equation}
Thus $k_{\rm GR}$ is the secular excess azimuthal phase accumulated per radial cycle, written in terms of the frequency variable used in the waveform evolution. The GR periastron-precession rate is
\begin{equation}
\dot{\gamma}_{\rm GR}=k_{\rm GR}\dot{l}=k_{\rm GR}\Omega_{r}.
\label{eq:gammadot-gr}
\end{equation}
Although the geometrical location of periastron is not defined for an
exactly circular orbit, the conservative dynamics has a smooth circular
limit. In this limit, $\Omega_r$ is interpreted as the radial epicyclic
frequency, while $\Omega_\theta = \dot{\theta}$ is the azimuthal frequency. Taking $e\rightarrow0$ in
Eq.~\eqref{eq:kgr} gives, at leading 1PN order~\cite{Blanchet:2002mb},
\begin{equation}
    K_{\rm GR}^{\rm circ} = \frac{\Omega_{\theta}}{\Omega_r} = 1+3x+\mathcal{O}(x^2),
    \label{eq:K-circular}
\end{equation}
where $x\equiv(M\Omega_{\theta})^{2/3}$. The difference between using $\Omega_\theta$ and $\Omega_r$ in the
definition of $x$ contributes only beyond 1PN order.

Our parametrization replaces
$K_{\rm GR}$ by $K_\alpha=1+(1+\delta\alpha)k_{\rm GR}$. Therefore, even in the circular limit, changing $K$ produces a relative 1PN modification of the conservative azimuthal phase evolution. However, since we use the eccentricity $e$ as a tracking parameter, the map between our modification and more usual 1PN expressions for circular orbits is not straightforward. The dissipative orbital evolution, by contrast, is kept fixed to its GR prediction.

The dissipative inspiral is governed by the 1PN orbit-averaged radiation-reaction equations,
\begin{equation}
\dot{\nu}=\dot{\nu}_{N}+\dot{\nu}_{\rm 1PN}, \qquad \dot{e}=\dot{e}_{N}+\dot{e}_{\rm 1PN}.
\label{eq:nudot-edot-1pn}
\end{equation}
Equivalently, the orbital-frequency evolution can be written using eccentricity as the independent variable,
\begin{equation}
\frac{d\nu}{de}=\frac{\dot{\nu}_{N}}{\dot{e}_{N}}+\frac{\dot{\nu}_{\rm 1PN}}{\dot{e}_{N}}-\frac{\dot{\nu}_{N}\dot{e}_{\rm 1PN}}{\dot{e}_{N}^{2}} .
\label{eq:dnu-de-1pn}
\end{equation}
This relation determines the 1PN function $\nu(e)$ used in the waveform construction. Solving Eq.~\eqref{eq:dnu-de-1pn} perturbatively gives the radial orbital frequency as a function of the radial eccentricity,
\begin{equation}
\nu(e)=\nu_{0}\frac{\sigma(e)}{\sigma(e_{0})}\left[1+\frac{3}{2}\nu_{0}^\frac{2}{3}\{b_{\rm PN}(e_{0})-\left(\frac{\sigma(e)}{\sigma(e_{0})}\right)^\frac{2}{3}b_{\rm PN}(e)\}\right],
\label{eq:nu-of-e-1pn}
\end{equation}
where
\begin{equation}
\sigma(e)=e^{-18/19}(1-e^{2})^{3/2}\left(1+\frac{121}{304}e^{2}\right)^{-1305/2299}.
\label{eq:sigma-e}
\end{equation}
Here $b_{\rm PN}(e)$ denotes the 1PN correction function given in Ref.~\cite{1PN_waveform}. Equation~\eqref{eq:nu-of-e-1pn} is the perturbative solution of Eq.~\eqref{eq:dnu-de-1pn} satisfying the boundary condition $\nu(e_{0})=\nu_{0}$.

Using eccentricity as the evolution variable, the time, radial phase, and GR periastron phase are obtained from
\begin{align}
t(e)-t_{0} &= \int_{e_{0}}^{e}\frac{d e'}{\dot{e}(e')},
\label{eq:time-of-e}
\\
l(e)- l_{0} &= \int_{e_{0}}^{e}\frac{\Omega_{r}(e')}{\dot{e}(e')}d e',
\label{eq:ell-of-e}
\\
\gamma_{\rm GR}(e)-\gamma_{0} &= \int_{e_{0}}^{e}\frac{\dot{\gamma}_{\rm GR}(e')}{\dot{e}(e')}d e' .
\label{eq:gamma-gr-of-e}
\end{align}
These functions define the GR inspiral on which the parametrized conservative-sector deformation is built.

\subsection{Time-domain eccentric waveform}
\label{subsec:time-domain-waveform}
The 1PN eccentric waveform contains Newtonian, half-PN, and 1PN amplitude contributions. In the source frame, the complex strain can be decomposed into spin-weighted spherical-harmonic modes,
\begin{equation}
h(t)\equiv h_{+}(t)-i h_{\times}(t) = \sum_{\ell,m}h_{\ell m}(t)\,
{}_{-2}Y_{\ell m}(\iota,\phi),
\label{eq:time-domain-mode-decomposition}
\end{equation}
where ${}_{-2}Y_{\ell m}(\iota,\phi)$ are the spin-weighted spherical
harmonics. Each mode can be written schematically as,
\begin{equation}
h_{\ell m}(t) = h_{\ell m}^{\rm N}(t) + h_{\ell m}^{\rm 0.5PN}(t) + h_{\ell m}^{\rm 1PN}(t).
\label{eq:pn-amplitude-decomposition}
\end{equation}
The explicit expressions for these terms are taken from the 1PN eccentric waveform of Ref.~\cite{1PN_waveform}.

For the waveform construction, it is useful to distinguish the part of the angular motion that is periodic over one radial cycle from the secular advance of the orbital orientation. The radial motion is
described by the mean anomaly $l(t)$, while the orientation of the ellipse in the orbital plane advances because of relativistic periastron precession. Schematically, the angular motion can be written as 
\begin{equation}
\theta(t) \simeq \varphi(t)+\gamma(t),
\label{eq:theta-sec-general}
\end{equation}
where $\varphi(t)$ describes the periodic radial motion and $\gamma(t)$ is the accumulated secular periastron-precession phase. The angular decomposition is schematic: at 1PN order, the orbital azimuth also contains a periodic angular
correction (see Eq.~\eqref{eq:theta-motion}), so that $\theta-\theta_0$ is not exactly equal to
$\varphi+\gamma$.

For any waveform contribution to which an angular phase with integer harmonic index $p$ is assigned, a generic factor ${\cal C}_{\ell m p}(e,\varphi) \exp[-ip\theta(t)]$ can be expanded in terms of $\varphi(t)$ and $\gamma(t)$ using Eq.~\eqref{eq:theta-motion} or Eq.~\eqref{eq:theta-sec-general},
where ${\cal C}_{\ell m p}$ is the amplitude function which is periodic in the radial motion. The
product of the amplitude function and the periodic
true-anomaly dependence can then be expanded in harmonics of the mean
anomaly,
\begin{equation}
{\cal C}_{\ell m p}(e,\varphi) \exp[-ip\varphi(t)] = \sum_n {\cal A}_{\ell mnp}(e(t)) \exp[-in l(t)].
\label{eq:hansen-expansion-schematic}
\end{equation}
This Fourier expansion is expressed in terms of generalized Hansen coefficients~\cite{1PN_waveform}.

A contribution with a specified secular-phase can therefore be written in the form
\begin{equation}
h_{\ell m}^{(p)}(t) = \sum_n {\cal A}_{\ell mnp}(e(t))
\exp\left\{-i\left[n l(t)+p\gamma(t)\right]\right\}.
\label{eq:time-domain-sideband-general}
\end{equation}
However, the set of waveform contributions to which this secular phase is assigned is model dependent. This assignment distinguishes the two waveform representations considered below: the
{\it restricted model} follows the explicit drift structure of the 1PN waveform in Ref.~\cite{1PN_waveform}, whereas the {\it full model}
promotes the secular phase relation to the lower-order angular carriers.

For a component carrying sideband index $p$, the corresponding GR phase is
\begin{equation}
\Phi_{np}^{\rm GR}(t) = n l(t)+p\gamma_{\rm GR}(t).
\label{eq:phase-np-gr}
\end{equation}

The representation in Eqs.~\eqref{eq:hansen-expansion-schematic}--\eqref{eq:phase-np-gr} makes the radial-harmonic index $n$ and the secular-precession label $p$ explicit. It provides a physically transparent sideband structure of the precession-dependent angular structures appearing in Ref.~\cite{1PN_waveform}. In that work, the periastron drift is introduced through the drift
anomaly, defined as $\varphi' = K_{\rm GR}\varphi = \left(1+k_{\rm GR}\right)\varphi$, and absorbed into the generally noninteger angular order of the
generalized Hansen coefficients. The subsequent expansion in the mean anomaly is a Fourier series and therefore contains only integer mean-anomaly harmonics by construction. The precession dependence is
encoded in the generally noninteger angular order and hence in the
harmonic coefficients, rather than appearing as a separate factor
$p\gamma_{\rm GR}$. We retain the explicit $(n,p)$ notation in the main text to identify the physical radial and precessional content and to specify which angular carriers receive the parametrized
deformation. The corresponding generalized-Hansen realization is described in Appendix~\ref{app:hansen-mapping}.

We now introduce the parametrized deformation of the conservative
precession relation and specify how it is assigned in the two waveform prescriptions.

\subsection{Parametrized periastron-precession deformation}
\label{subsec:parametrized-precession}

We introduce a dimensionless phenomenological non-GR parameter
$\delta\alpha$ that deforms the GR prediction for the conservative
periastron advance,
\begin{equation}
k_{\alpha}(e,\nu) = \left(1+\delta\alpha\right)k_{\rm GR}(e,\nu).
\label{eq:k-alpha}
\end{equation}
Equivalently,
\begin{equation}
K_{\alpha} = 1+k_{\alpha} = 1+\left(1+\delta\alpha\right)k_{\rm GR}.
\label{eq:capital-k-alpha}
\end{equation}
The corresponding deformed precession rate is
\begin{equation}
\dot{\gamma}_{\alpha} = k_{\alpha}\dot{l} = \left(1+\delta\alpha\right)\dot{\gamma}_{\rm GR}.
\label{eq:gammadot-alpha}
\end{equation}
Positive values of $\delta\alpha$ correspond to faster periastron
precession than predicted by GR, while negative values correspond to
slower precession.

We define the accumulated deformed periastron phase by keeping the
initial periastron angle $\gamma_0$ fixed and modifying only the
precession accumulated after the start of the observation,
\begin{equation}
\gamma_{\alpha}(e) = \gamma_0 + \left(1+\delta\alpha\right)\left[
\gamma_{\rm GR}(e)-\gamma_{\rm GR}(e_0)
\right].
\label{eq:gamma-alpha}
\end{equation}
This convention ensures that $\delta\alpha$ changes the accumulated relativistic precession rather than shifting the arbitrary initial
orientation $\gamma_0$.

For any waveform component to which the secular precession phase is
assigned, the corresponding deformed phase is
\begin{equation}
\Phi_{np}(t;\delta\alpha) = n l(t)+p\gamma_{\alpha}(t).
\label{eq:phase-np-alpha}
\end{equation}

The deformation in Eqs.~\eqref{eq:k-alpha}--\eqref{eq:gamma-alpha} is applied only to the conservative precession sector. The radiation-reaction functions $\dot{\nu}$ and $\dot{e}$, and hence the relation $\nu(e)$, are kept fixed to their 1PN GR values. The radiative amplitude coefficients are also kept fixed to their GR values. The parameter $\delta\alpha$
therefore isolates a deviation in the azimuthal--radial frequency ratio rather than a modification of the dissipative dynamics or of the source multipole amplitudes.

A subtlety arises when the dynamical deformation $K_{\rm GR}\to K_{\alpha}$ is embedded in the waveform. Reference~\cite{1PN_waveform} organizes the secular periastron
precession through the drift anomaly and retains explicit
precession-phase dependence only in a subset of the 1PN harmonic
structures. We adopt this organization as the baseline for our first representation. Alternatively, the same secular azimuthal--radial phase
relation can be assigned to the lower-order angular carriers before the harmonic decomposition. Retaining this phase without re-expansion
produces a secularly resummed waveform and therefore defines a distinct finite approximant.

We consequently introduce two prescriptions. In the restricted model, denoted by $\mathcal{H}_{\rm R}$, the replacement $K_{\rm GR}\to K_{\alpha}$ is applied only to the
explicit precession-dependent 1PN structures. Setting $\delta\alpha=0$ therefore recovers the GR waveform in Ref.~\cite{1PN_waveform}. In the full model, denoted by $\mathcal{H}_{\rm F}$, the corresponding secular phase is also assigned to the lower-order angular carriers. Its $\delta\alpha=0$ limit is therefore a secularly resummed GR waveform and is not, in general, identical to the GR waveform in Ref.~\cite{1PN_waveform}. Both prescriptions deform the same conservative frequency ratio and introduce no additional dynamical degree of freedom; they differ only in which radiative carriers inherit the secular precession phase before the frequency-domain waveform is constructed.

\subsubsection{Restricted model}
\label{subsubsec:restricted-model}
In the restricted implementation, the deformation is introduced through the explicit periastron-precession sidebands of the  1PN waveform. The source-frame spherical-harmonic modes are written
schematically as
\begin{align}
h_{\ell m}^{\rm R}(t;\delta\alpha) {}& =
h_{\ell m}^{\rm N,GR}(t) + h_{\ell m}^{\rm 0.5PN,GR}(t)
\nonumber\\
&+
\sum_n
\sum_{p\in{\cal P}_{\rm R}}
{\cal A}_{\ell mnp}^{\rm GR,1PN}(e(t))
\exp\left[-i\Phi_{np}(t;\delta\alpha)\right],
\label{eq:time-domain-sb}
\end{align}
where $n$ labels the radial eccentric harmonic, $p$ labels the
explicit precession sideband, and $\Phi_{np}(t;\delta\alpha)$ is
defined by Eq.~\eqref{eq:phase-np-alpha}. The Newtonian and half-PN
contributions retain their GR values, while
the GR 1PN amplitude coefficients ${\cal A}_{\ell mnp}^{\rm GR,1PN}$ are not directly deformed.

In this implementation, the explicit 1PN sideband set is
\begin{equation}
{\cal P}_{\rm R} = \{-4,-2,0,2,4\}.
\label{eq:p-values-sb}
\end{equation}
For $p=0$, the factor multiplying $\gamma_\alpha$ vanishes, so these terms terms are independent of $\delta\alpha$ in this restricted implementation, whereas the $p\neq0$ terms contain $\gamma_\alpha$ and are directly sensitive to the deformation. This implementation is conservative in the sense that $\delta\alpha$ is restricted to the explicit precession-dependent structure. At $\delta\alpha=0$, it therefore reduces to the GR waveform in Ref.~\cite{1PN_waveform}.

\subsubsection{Full model}
\label{subsubsec:carrier-model}
The full implementation is motivated by the physical interpretation of periastron advance as a deformation of the secular azimuthal--radial frequency ratio,
$K= \Omega_\theta/ \Omega_r.$ We therefore promote the replacement $K_{\rm GR}\to K_{\alpha}$ to the angular
carriers entering the Newtonian, half-PN, and 1PN waveform pieces. At fixed orbital elements, the secular angular relation is obtained by replacing
$\theta_{\rm GR} = \left(1+k_{\rm GR}\right)\varphi$ by
\begin{equation}
\theta_{\rm F} = \left[ 1+\left(1+\delta\alpha\right)k_{\rm GR}
\right]\varphi.
\label{eq:carrier-replacement}
\end{equation}
The periodic 1PN contribution $\kappa_2\sin\varphi$ in
Eq.~\eqref{eq:theta-motion} is not independently deformed and is
retained at its GR value. Equivalently, the angular phase can be written as
\begin{align}
\theta_{\rm F}(t) &\simeq \varphi(t)+\gamma_{\alpha}(t),
\label{eq:theta-gamma-connection}\\
\dot{\theta}_{\rm F} & \simeq \Omega_r+\dot{\gamma}_{\alpha} = K_{\alpha}\Omega_r.
\label{eq:theta-dot-cp}
\end{align}
These equations make explicit how the deformation of $K$ enters the
waveform phase. The carrier-phase-deformed modes can be written as
\begin{align}
{}& h_{\ell m}^{\rm F}(t;\delta\alpha)
 =
\sum_n\sum_{p=\pm2} {\cal A}_{\ell mnp}^{\rm GR,N}(e(t))
\exp\left[-i\Phi_{np}(t;\delta\alpha)\right]
\nonumber\\
&
 \;\;\; \;\;\; + \sum_n\sum_{p=\pm1,\pm3}
{\cal A}_{\ell mnp}^{\rm GR,0.5PN}(e(t))
\exp\left[-i\Phi_{np}(t;\delta\alpha)\right]
\nonumber\\
&
 \;\;\; \;\;\; + \sum_n\sum_{p=0,\pm2,\pm4}
{\cal A}_{\ell mnp}^{\rm GR,1PN}(e(t))
\exp\left[-i\Phi_{np}(t;\delta\alpha)\right].
\label{eq:time-domain-cp}
\end{align}
The difference from the restricted representation is the assignment of the secular phase
$\gamma_{\alpha}$ to the lower-order angular carriers. The full retained sideband set is therefore
\begin{equation}
{\cal P}_{\rm F} = \{-4,-3,-2,-1,0,1,2,3,4\}.
\label{eq:p-values-cp}
\end{equation}
The origin of this enlarged set can be traced to the angular carriers
at each PN amplitude order. The Newtonian quadrupole carriers
contribute $p=\pm2$, the half-PN carriers contribute
$p=\pm1,\pm3$, and the explicit 1PN terms contribute $p=0,\pm2,\pm4$. Thus, the labels $p=\pm2$ occur in both representations, but in $\mathcal{H}_{\rm F}$ they are also attached to the louder Newtonian quadrupole carriers. The odd sidebands arise from the promotion of the secular phase to the half-PN angular carriers. Because the Newtonian and half-PN carriers contain a larger fraction
of the signal power than the explicit 1PN sidebands, the
full model retains sensitivity to $\delta\alpha$
through the accumulated phase of the dominant waveform components,
rather than only through amplitude-suppressed explicit 1PN precession sidebands.

Setting $\delta\alpha=0$ restores the GR frequency ratio,
$K_\alpha=K_{\rm GR}$, in both prescriptions, so they describe the
same underlying GR conservative dynamics. They do not, however,
reduce to the same finite waveform approximant. In
Ref.~\cite{1PN_waveform}, the underlying orbital dynamics includes
the GR periastron advance, while its appearance in the waveform is
organized according to PN order. The Newtonian and half-PN waveform
contributions are expressed in harmonics of the generalized true
anomaly $\varphi$, whereas the explicit secular drift-anomaly
dependence, $\varphi'=(1+k_{\rm GR})\varphi$
appears in the 1PN contribution. The waveform of
Ref.~\cite{1PN_waveform} should therefore not be interpreted as
assigning a factor $\exp[-ip\gamma(t)]$, universally to the Newtonian, half-PN, and 1PN angular carriers.

Our restricted model follows this organization and applies the drift
phase only to the explicit precession-dependent structures appearing
in the 1PN waveform contribution. The full model instead assigns the
same secular drift phase also to carriers with Newtonian and half-PN
amplitudes. Since $k_{\rm GR}=O(v^2)$, including this phase in a
Newtonian-amplitude carrier modifies the waveform beginning at
relative 1PN order, while including it in a half-PN-amplitude carrier
generates corrections beginning at relative 1.5PN order. The latter
lies beyond the formal 1PN truncation of Ref.~\cite{1PN_waveform}.
Because the secular phase is retained without re-expansion in the full
model, these corrections are kept in resummed form. Consequently,
even at $\delta\alpha=0$, the full and restricted prescriptions remain
different finite waveform approximants.

The distinction between $\mathcal{H}_{\rm R}$ and
$\mathcal{H}_{\rm F}$ therefore concerns how the same conservative
frequency ratio is incorporated into the waveform, rather than the
dynamical quantity being tested. The full construction may be
interpreted as evaluating the retained radiative multipoles along the
secularly precessing azimuthal trajectory and should be regarded as a
secularly resummed phenomenological waveform based on the same 1PN
conservative precession relation, rather than as a strict
reorganization of the finite 1PN waveform of
Ref.~\cite{1PN_waveform}. We use the full model as our fiducial
phenomenological prescription, while retaining the restricted model
as a controlled baseline that remains closest to the waveform
organization of Ref.~\cite{1PN_waveform}.

\subsection{Frequency-domain waveform}
\label{subsec:sideband-spa}
For physical interpretation, we express both
$\mathcal{H}_{\rm R}$ and $\mathcal{H}_{\rm F}$ in a resolved
two-frequency representation. Rewritting Eq.~\eqref{eq:phase-np-alpha}, a component with radial-harmonic index $n$ and precession label $p$ carries the phase
\begin{equation}
\Phi_{np}(t;\delta\alpha)=n l(t)+p\gamma_{\alpha}(t),
\label{eq:spa-time-phase}
\end{equation}
We construct the corresponding frequency-domain contribution using the
stationary-phase approximation~\cite{Cutler:1994ys}. For the Fourier convention adopted here, the stationary time $t_{np}(f)$ is determined by
\begin{equation}
2\pi f=\frac{d\Phi_{np}}{dt}=n\dot{l}+p\dot{\gamma}_{\alpha}.
\label{eq:spa-condition}
\end{equation}
Using $\dot{l}=2\pi\nu$, this defines the sideband frequency map
\begin{equation}
f=F_{np}(e;\delta\alpha)=n\nu(e)+\frac{p}{2\pi}\dot{\gamma}_{\alpha}(e).
\label{eq:sideband-frequency-map}
\end{equation}
For a given Fourier frequency $f$, the saddle-point eccentricity $e_{np}(f;\delta\alpha)$ is obtained implicitly from
$f=F_{np}\left(e_{np};\delta\alpha\right).$
The corresponding SPA phase is
\begin{equation}
\Psi_{np}(f;\delta\alpha)=2\pi f t(e_{np})-n l(e_{np})-p\gamma_{\alpha}(e_{np})-\frac{\pi}{4},
\label{eq:spa-phase}
\end{equation}
where $e_{np}=e_{np}(f;\delta\alpha)$. The deformation parameter enters the SPA phase both explicitly through $\gamma_{\alpha}$ and implicitly through the saddle point $e_{np}(f;\delta\alpha)$.

The source-frame spherical-harmonic mode is
\begin{equation}
\tilde{h}_{\ell mnp}^{X}(f;\delta\alpha)=\frac{{\cal A}_{\ell mnp}^{\rm GR,X}\left(e_{np}\right)}{\sqrt{\dot{F}_{np}\left(e_{np};\delta\alpha\right)}}\exp\left[i\Psi_{np}(f;\delta\alpha)\right],
\label{eq:single-sideband-spa}
\end{equation}
where $X\in\{{\rm R},{\rm F}\}$ labels the waveform implementation and
$\dot{F}_{np}(e;\delta\alpha)= (dF_{np}/{de})\dot{e}(e)$.
We do not introduce an independent amplitude deformation. Nevertheless, the sideband amplitude can acquire an implicit $\delta\alpha$-dependence through the saddle eccentricity $e_{np}(f;\delta\alpha)$ and the SPA prefactor $\dot{F}_{np}^{-1/2}$.

The full source-frame spherical-harmonic mode is obtained by summing over all retained radial harmonics and sidebands,
\begin{equation}
\tilde{h}_{\ell m}^{X}(f;\delta\alpha)=\sum_{n}\sum_{p\in{\cal P}_{X}}\tilde{h}_{\ell mnp}^{X}(f;\delta\alpha).
\label{eq:mode-sum}
\end{equation}
Equations~\eqref{eq:spa-time-phase}--\eqref{eq:mode-sum} provide a physically transparent resolved-sideband representation of the waveform. In the numerical
realization, the drifted true-anomaly factors are evaluated using the
generalized-Hansen construction of Ref.~\cite{1PN_waveform}. In that
basis, the precession label $p$ and the secular phase
$\gamma_{\alpha}$ are absorbed into generally noninteger angular
orders, and the resulting factors are projected onto integer harmonics
of the mean anomaly. The explicit $(n,p)$ notation is retained here to
identify the physical radial and precessional content of each
contribution, and, in particular, to specify which angular carriers
receive the deformation. The generalized-Hansen representation is a
computational reorganization of this drift-dependent angular structure;
its harmonic labels need not correspond one-to-one to the resolved
$(n,p)$ labels. Further details are given in
Appendix~\ref{app:hansen-mapping}.

The complex strain in the source frame is reconstructed as
\begin{align}
\tilde{h}^{X}(f;\delta\alpha) &= \tilde{h}_{+}^{X}(f;\delta\alpha)-i\tilde{h}_{\times}^{X}(f;\delta\alpha)
\nonumber\\
&= \sum_{\ell,m}\tilde{h}_{\ell m}^{X}(f;\delta\alpha){}_{-2}Y_{\ell m}(\iota,\phi),
\label{eq:source-strain}
\end{align}
The source-frame modes in Eq.~\eqref{eq:mode-sum} are passed through the time-dependent LISA response for each retained component $a\equiv(\ell,m,n,p)$.

\subsection{LISA response and likelihood}
\label{subsec:lisa-response}
We use ~\textsc{lisabeta}~\cite{PhysRevD.103.083011} for performing the parameter estimation. \textsc{lisabeta} evaluates the corresponding frequency-domain TDI response $\tilde{h}_{I,a}^{\rm TDI}(f;\delta\alpha),  I\in\{A,E,T\},$ including the frequency-dependent transfer functions and the motion of the LISA constellation. The total TDI template is obtained by summing the response of all retained components before evaluating the likelihood,
\begin{equation}
\tilde{h}_{I,{\rm tot}}^{\rm TDI}(f;\delta\alpha)=\sum_{a}\tilde{h}_{I,a}^{\rm TDI}(f;\delta\alpha).
\label{eq:tdi-sum}
\end{equation}
The response of each retained eccentric harmonic $n$ is evaluated on
its own frequency grid over the corresponding harmonic support, with
the resolution chosen to resolve its time--frequency evolution over
the observation. For the likelihood evaluation, the individual TDI
contributions are combined on a common frequency grid constructed from the union of the harmonic grids and summed before computing the inner
product. This construction retains interference between different
eccentric harmonics $n$, precession sidebands $p$, and
spherical-harmonic modes $(\ell,m)$, rather than approximating the
inner product by a diagonal sum over waveform components. We validated the numerical accuracy of this procedure by comparing the
production configuration with more densely sampled reference
waveforms. The resulting total-waveform mismatches remain below the
parameter-estimation criterion $1/(2\rho^2)$ at the highest SNRs
considered, and the likelihood is stable under further refinement of
the frequency grids. In particular,
\begin{equation}
\left(h_{\rm tot}^{\rm TDI}\middle|h_{\rm tot}^{\rm TDI}\right)=\sum_{a,b}\left(h_{a}^{\rm TDI}\middle|h_{b}^{\rm TDI}\right),
\label{eq:cross-terms}
\end{equation}
so diagonal and cross terms are included automatically without evaluating all pairwise overlaps explicitly.

In Bayesian inference~\cite{Veitch:2009hd, 2019PASA...36...10T}, the posterior probability distribution on waveform parameters $\boldsymbol{\theta}$ given data $d$ from GW observations can be computed using Bayes’ theorem.
\begin{equation}
    p(\boldsymbol{\theta} | d) \propto \mathcal{L}(d | \boldsymbol{\theta}) \pi(\boldsymbol{\theta}) \,,
\end{equation}
where $\mathcal{L}(d | \boldsymbol{\theta})$ is the likelihood function of data given $\boldsymbol{\theta}$ and $\pi(\boldsymbol{\theta})$ is the prior probability for $\boldsymbol{\theta}$. Assuming stationary Gaussian noise, the likelihood function can be written as
\begin{equation}
{\cal L}(d|\boldsymbol{\theta})\propto \exp\left[-\frac{1}{2}\left(d-h_{\rm tot}^{\rm TDI}(\boldsymbol{\theta})\middle|d-h_{\rm tot}^{\rm TDI}(\boldsymbol{\theta})\right)\right].
\label{eq:likelihood}
\end{equation}
where the noise-weighted inner product between two signals is defined as
\begin{equation}
    (a|b) = 4\,{\rm Re} \sum_{I\in\{A,E,T\}} \int_{f_{\rm low}}^{f_{\rm high}}
    \frac{
        \widetilde{a}_{I}(f)
        \widetilde{b}_{I}^{*}(f) }{
        S_{I}(f) } \,df ,
    \label{eq:inner-product}
\end{equation}
where $S_I(f)$ is the one-sided noise power spectral density of the TDI channel $I$, and $*$ denotes complex conjugation. We adopt the LISA noise model provided by the LISA Science Requirements document~\cite{LISAsr:18aa}. 

The parameter vector sampled in the Bayesian analysis is
\begin{equation}
    \boldsymbol{\theta}= \{e_0,f_0,\mathcal{M},q,d_L,\iota,\phi,
    \lambda,\beta,\psi,\gamma_0, l_0,\delta\alpha
    \}.
    \label{eq:parameter-vector}
\end{equation}
Here $e_0$ and $f_0=2\nu_0$ are, respectively, the initial eccentricity and
the quadrupole gravitational-wave frequency at the beginning of the
observation. The remaining parameters are the detector-frame chirp
mass $\mathcal{M}$, the mass ratio $q=m_1/m_2$, the luminosity
distance $d_L$, the inclination $\iota$, the observer azimuth $\phi$,
the sky position $(\lambda,\beta)$, the polarization angle $\psi$,
the initial periastron phase $\gamma_0$, and the initial mean anomaly
$\ell_0$.

For each injection, we specify the eccentricity $e_0$ at the beginning
of the LISA observation and determine the initial radial orbital
frequency $\nu_0$ by requiring the binary to evolve for the prescribed
observation duration $T_{\rm obs}$ before reaching the adopted endpoint
of the inspiral. Using the 1PN frequency--eccentricity relation
$\nu(e;\nu_0,e_0)$ obtained from
Eq.~\eqref{eq:dnu-de-1pn}, this condition is
\begin{equation}
    t_c =\int_{e_{\rm end}}^{e_0} \frac{de}{
        -\dot e\!\left[e,\nu(e;\nu_0,e_0)\right]},
    \label{eq:solve-nu0-from-tobs}
\end{equation}
where $\dot e<0$ during the inspiral. 

The $n$th radial harmonic is retained only over the frequency interval
$f_{\rm low}^{(n)} \leq f \leq f_{\rm high}^{(n)}$. The lower cutoff is
\begin{equation}
    f_{\rm low}^{(n)} = \max\!\left[10^{-5}\,,\,n\nu_0\right] = \max\!\left[10^{-5}\,,\,\frac{n}{2}f_0\right],
    \label{eq:frequency-low}
\end{equation}
where $f_0=2\nu_0$ is the initial quadrupole GW frequency.

To impose the finite observation duration, we determine the eccentricity
$e_{\rm obs}$ reached after an elapsed time $T_{\rm obs}$ from the
initial point $(e_0,\nu_0)$ through
\begin{equation}
    T_{\rm obs} = \int_{e_{\rm obs}}^{e_0} \frac{de} {-\dot e[e,\nu(e;\nu_0,e_0)]},
    \label{eq:eobs}
\end{equation}
provided that the binary does not reach the adopted inspiral endpoint
earlier. The corresponding radial orbital frequency is
\begin{equation}
    \nu_{\rm obs} = \nu(e_{\rm obs};\nu_0,e_0).
\end{equation}
The upper frequency of the $n$th radial harmonic is then
\begin{equation}
    f_{\rm high}^{(n)} = \min\!\left[0.2,\,f_{\rm ISCO}^{(n)},\, n\nu_{\rm obs}
    \right], \;\;\; f_{\rm ISCO}^{(n)} =  \frac{n}{2\pi 6^{3/2}M}.
    \label{eq:frequency-high}
\end{equation}
The lower cutoff prevents the $n$th harmonic from being evaluated before
the reference point $(e_0,\nu_0)$. The frequencies
$10^{-5}\,\mathrm{Hz}$ and $0.2\,\mathrm{Hz}$ define the adopted LISA
analysis band, $f_{\rm ISCO}^{(n)}$ imposes the harmonic-dependent
Schwarzschild-ISCO termination, and $n\nu_{\rm obs}$ truncates the
waveform at the end of the finite observation window.

The radial-harmonic sum is truncated at
$n_{\max}$~\cite{OLeary:2008myb},
\begin{equation}
    n_{\max}(e_0)= \left\lfloor 5\frac{\sqrt{1+e_0}}{(1-e_0)^{3/2}}
    \right\rfloor ,
    \label{eq:nmax-oleary}
\end{equation}
where $\lfloor\cdot\rfloor$ denotes the floor function. This
prescription was constructed to retain approximately $99\%$ of the
emitted signal power and therefore provides a conservative truncation criterion. It becomes overly conservative at low eccentricity, however, since it approaches $n_{\max}=5$ as $e_0\rightarrow0$, retaining harmonics beyond the dominant $n=2$ circular contribution.

An alternative estimate is the fit introduced by
Wen~\cite{Wen:2002km},
\begin{equation}
    n_{\rm peak}(e_0) = 2\frac{(1+e_0)^{1.1954}}{(1-e_0^2)^{3/2}},
    \label{eq:npeak-wen}
\end{equation}
which approaches the dominant quadrupole harmonic,
$n_{\rm peak}\rightarrow2$, in the circular limit. This quantity
estimates the harmonic at which the instantaneous emitted power is
maximal, but it is not designed to retain a specified fraction of the
total signal power.

We compare the two prescriptions in
Appendix~\ref{app:harmonic-convergence} with the detector-weighted
harmonic content of the present waveform.
Figure~\ref{fig:n99-eccentricity} shows that truncation at
$n_{\rm peak}$ does not retain $99\%$ of the cumulative diagonal
$(l,m)=(2,2)$ SNR squared, particularly for highly eccentric systems
observed several years before merger. By contrast,
Eq.~\eqref{eq:nmax-oleary} remains conservative over the investigated
range. We therefore adopt $n\leq n_{\max}(e_0)$ in Eq.~\eqref{eq:nmax-oleary} in the waveform generation.

\subsection{Spectral content of the eccentric frequency-domain waveform}

\begin{figure*}\label{fig:waveform}
    \centering
    \begin{subfigure}[]
  {\includegraphics[width=0.32\textwidth]{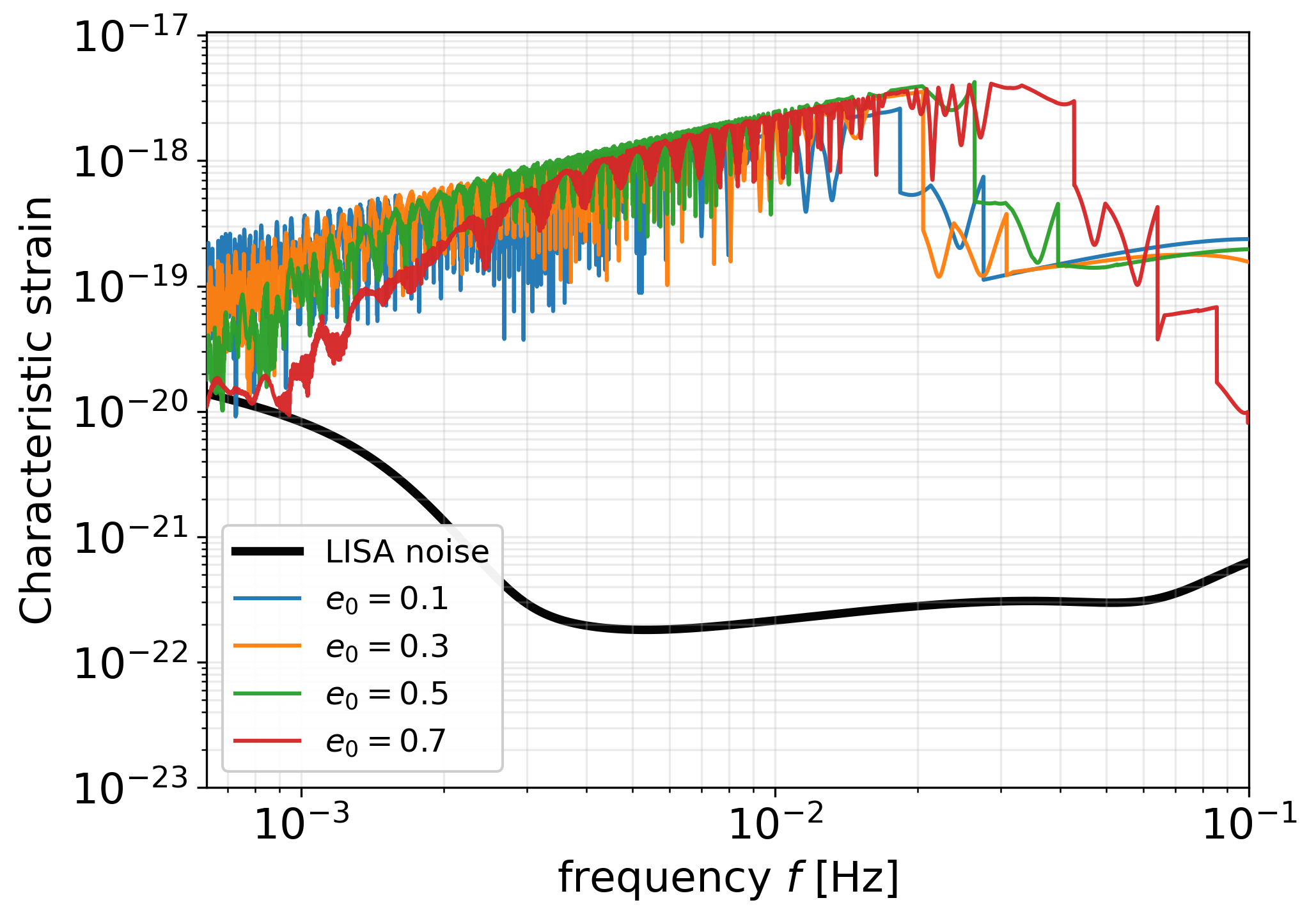}}\label{fig:strain_vs_e0}
  \end{subfigure}
  \begin{subfigure}[]
 {\includegraphics[width=0.32\textwidth]{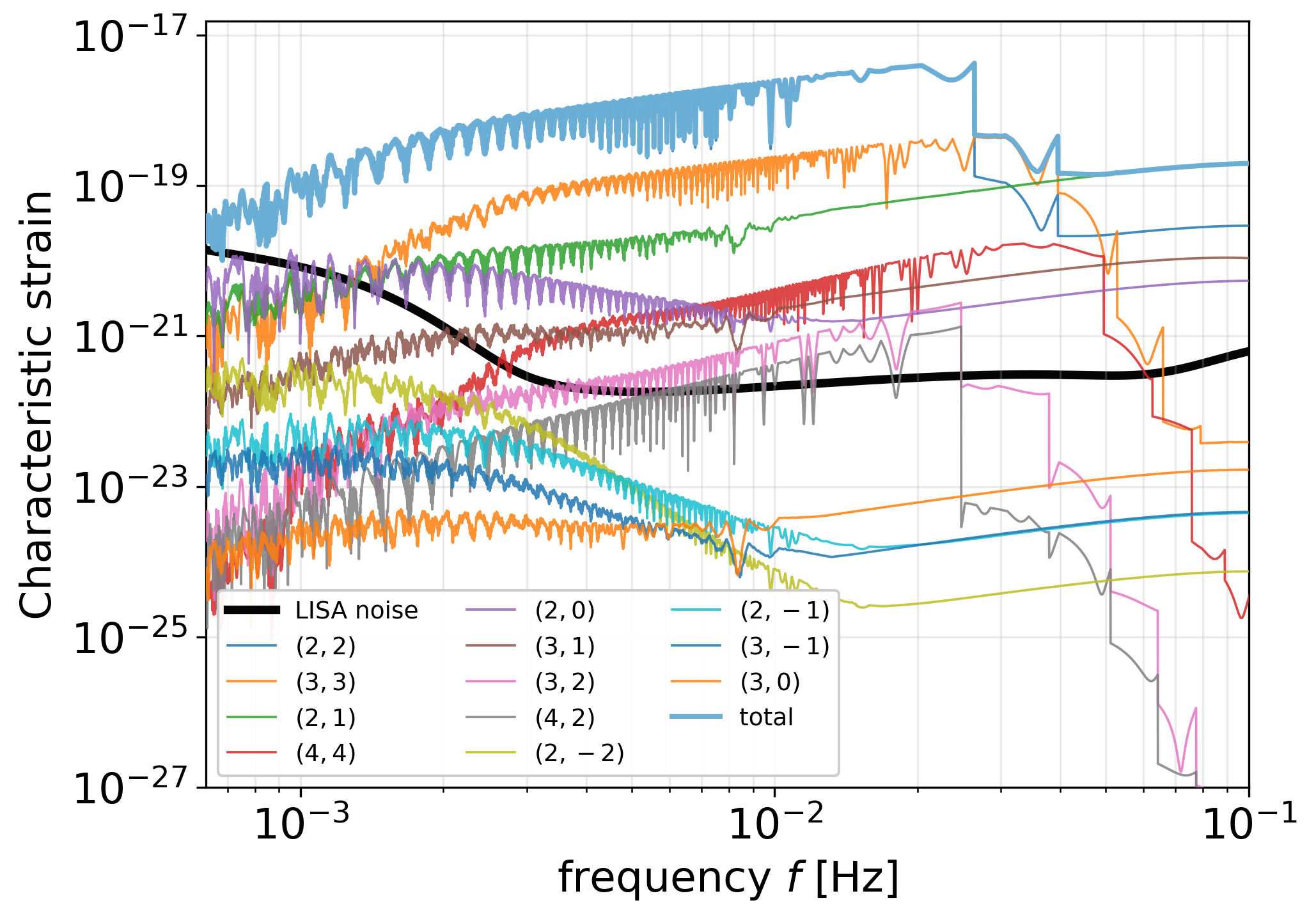}}\label{fig:hlm_modes}
  \end{subfigure}
\begin{subfigure}[]
 {\includegraphics[width=0.32\textwidth]{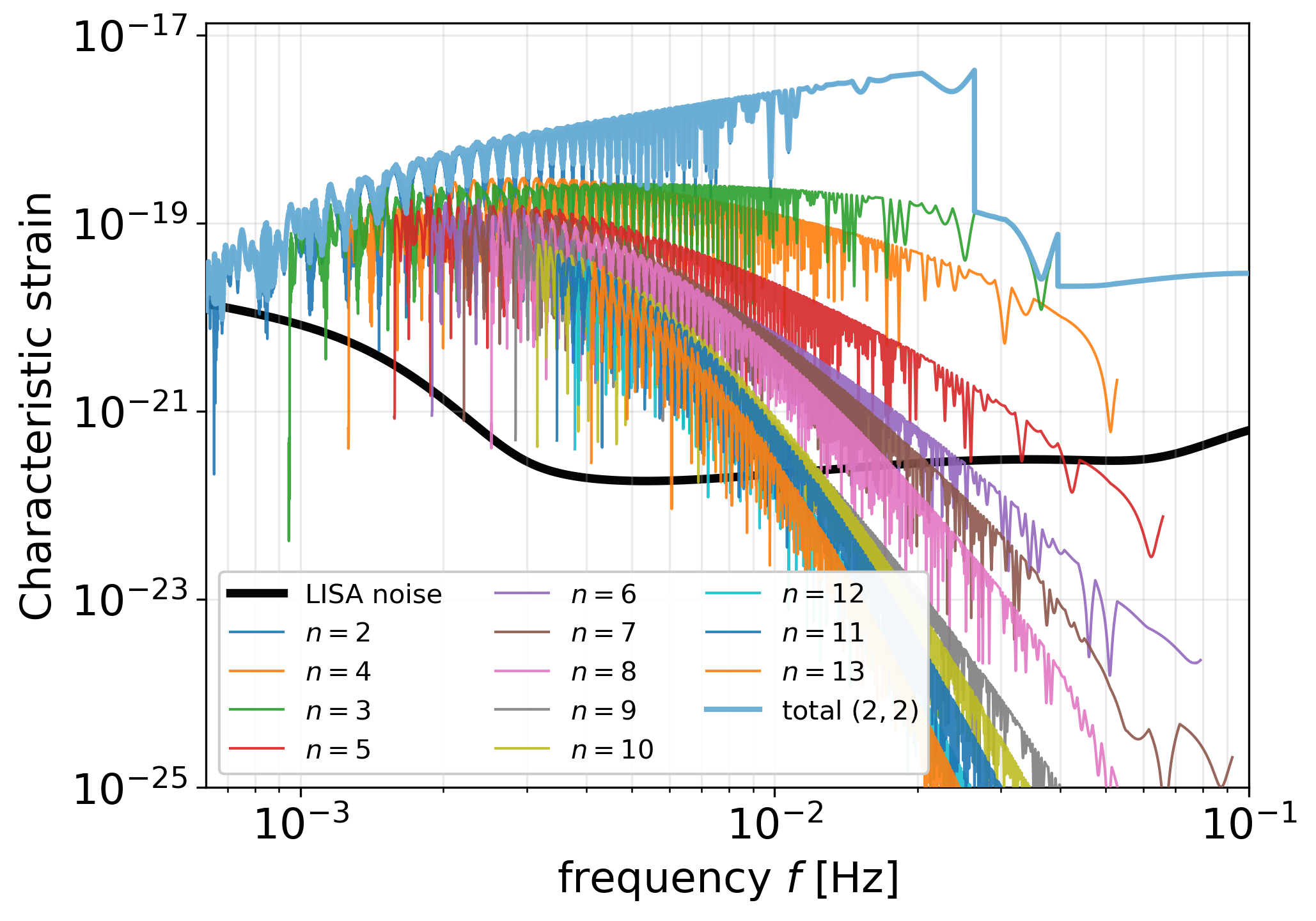}}\label{fig:h22_eccentric_harmonics}
  \end{subfigure}
   \caption{Characteristic strain of the frequency-domain eccentric waveform in the LISA TDI-A channel. The source has chirp mass $\mathcal{M}=3000\,M_{\odot}$, mass ratio $q=2$. The common starting frequency is set to $f_{0}=6.3\times10^{-4}\,{\rm Hz}$, obtained from the fiducial $e_0=0.5$ system by placing it $10\,{\rm yr}$ before merger. The black curve shows the LISA TDI-A characteristic noise. {\it Left}: total characteristic-strain envelope for different initial eccentricities, $e_0=\{0.1,0.3,0.5,0.7\}$, all started at the same $f_{0}$. {\it Middle}: decomposition of the $e_0=0.5$ signal into the dominant $(\ell,m)$ modes, where all eccentric harmonics and periastron sidebands are included for each mode. {\it Right}: decomposition of the $(2,2)$ mode into different eccentric harmonics $n$, with the periastron sidebands summed for each harmonic.
\label{fig:lisa_characteristic_strain_eccentric}}
\end{figure*}
To understand the content of the waveform, we decompose the waveform into different modes and plot their harmonic content. For a single Fourier-domain contribution, the characteristic strain is defined as 
\begin{equation}
    h_c(f) = 2 f |\tilde h(f)|\,.
\end{equation}
However, in the eccentric waveform, each mode family receives contributions from several eccentric harmonics $n$ and periastron sidebands $p$. To display the distribution of signal power more clearly, we define the envelope of a given $(\ell,m)$ mode as
\begin{equation}
    h^{\rm env,A}_{c,\ell m}(f) =  2f \left[\sum_{n,p}
    \left|  \tilde h^{A}_{\ell m n p}(f) \right|^2 \right]^{1/2},
    \label{eq:hc_lm_envelope}
\end{equation}
where $A$ denotes the LISA TDI-A channel. The corresponding total envelope is obtained by summing over all $(\ell, m)$ modes
\begin{equation}
    h^{\rm env,A}_{c,\rm tot}(f) = 2f \left[\sum_{\ell,m,n,p}
    \left| \tilde h^{A}_{\ell m n p}(f)  \right|^2 \right]^{1/2}.
    \label{eq:hc_total_envelope}
\end{equation}
Figure~\ref{fig:lisa_characteristic_strain_eccentric} illustrates the spectral content of the eccentric Fourier-domain waveform after applying the LISA TDI-A response. We consider a representative binary with  $\mathcal{M}=3000\,M_{\odot}$, and $q=2$. The initial frequency is fixed to a common value,
$f_{0}=6.3\times10^{-4}\,{\rm Hz}$, which is obtained by evolving the $e_0 = 0.5$ system  backwards to $10\,{\rm yr}$ before merger. This choice allows us to compare the spectral content of binaries with different eccentricities at the same reference frequency.

The left panel of Fig.~\ref{fig:lisa_characteristic_strain_eccentric} shows the total envelope for different initial eccentricities. 
Since all systems are started at the same $f_{0}$, the differences between the curves are driven by the eccentric redistribution of power. Increasing the eccentricity broadens the signal content and shifts power into higher harmonics, modifying the frequency dependence of the observable strain.

The middle panel decomposes the $e_0=0.5$ waveform into its dominant $(\ell,m)$ modes using Eq.~\eqref{eq:hc_lm_envelope}. The quadrupolar $(2,2)$ mode gives the largest contribution, but several subdominant modes remain visible in the LISA band. 
These additional modes contribute to the broadband structure of the eccentric signal and become important for accurately modelling the full response.

The right panel further decomposes the $(2,2)$ mode into individual eccentric harmonics. For each $n$, we plot
\begin{equation}
    h^{\rm env,A}_{c,22,n}(f) =2f \left[\sum_p\left|\tilde h^{A}_{22np}(f) \right|^2 \right]^{1/2}.
    \label{eq:hc_22n_envelope}
\end{equation}
This shows that the eccentric signal is not confined to the quasicircular $n=2$ harmonic. Instead, for moderate eccentricity, a broad set of eccentric harmonics contributes appreciably to the observed strain. This harmonic redistribution is one of the main spectral signatures of eccentricity in the LISA band.

\subsubsection{SNR contributions of different modes}
To understand which waveform components dominate the LISA signal, we first examine the SNR budget (in the GR limit, $\delta\alpha=0$). This is useful because the mode and harmonic content of the GR waveform determines where most of the statistical information in the signal resides before introducing the parametrized precession deformation. Since the two prescriptions retain the same underlying GR radiative amplitude content, their leading SNR budget is expected to be very similar. Small differences between the two finite waveform representations can nevertheless arise through their different phase organization and the associated interference terms. 

\begin{figure}
\centering
\includegraphics[width=0.92\columnwidth]{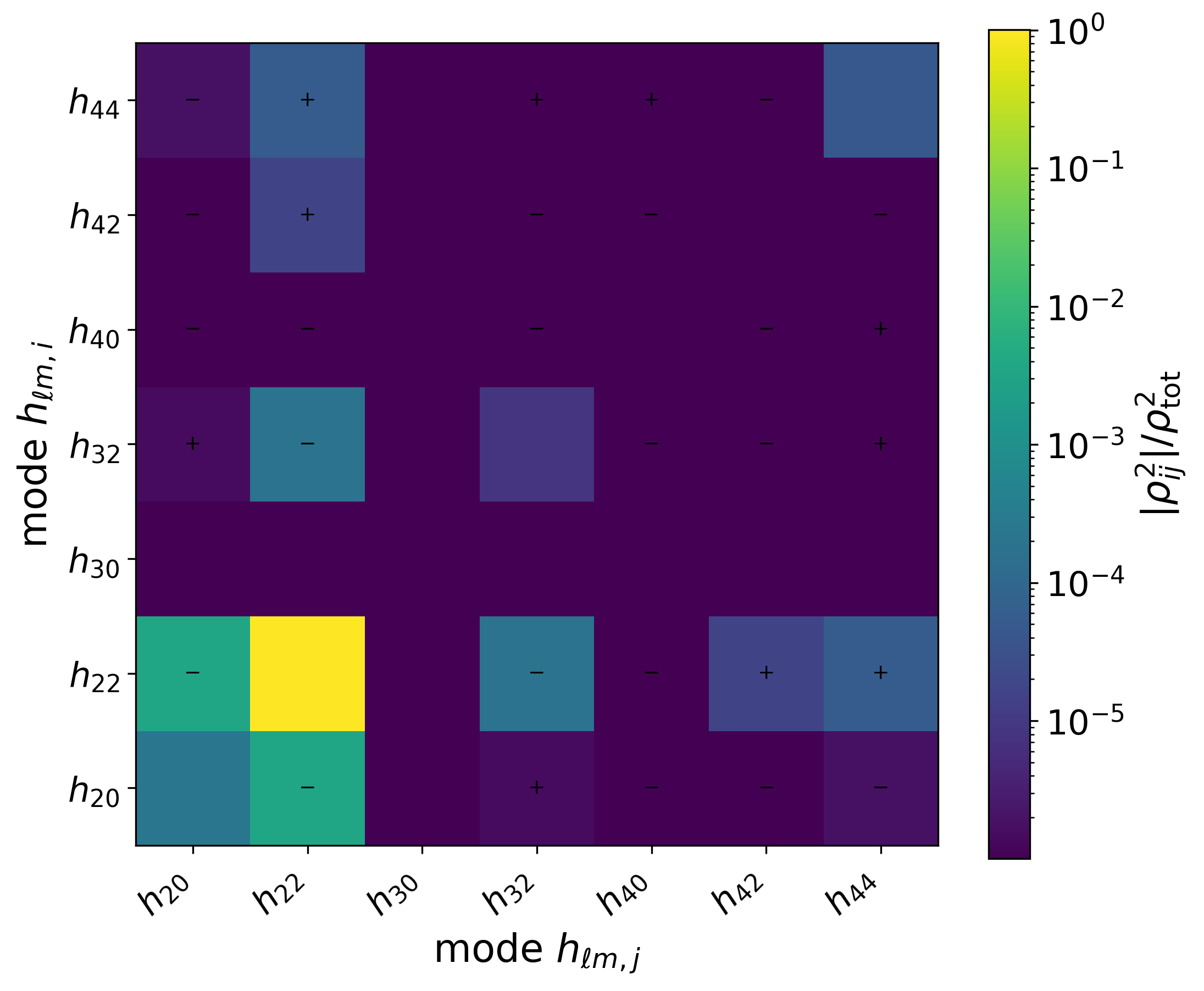}
\caption{Fractional SNR budget among source-frame modes. The $+$ and $-$ symbols indicate the sign of the off-diagonal cross terms. The source parameters are $e_{0}=0.4$, ${\cal M}=3000 M_{\odot}$, $q=5$, $\iota=\pi/2$, $T_{\rm obs}=4\, {\rm yr}$, and $t_{c}=4 \, {\rm yr}$. As expected, the SNR budget is dominated by the $(\ell,m)=(2,2)$ mode; higher-mode diagonal contributions and cross-mode contributions are subdominant.}
\label{fig:mode-snr-matrix}
\end{figure}

We first decompose the GR waveform into source-frame modes,
\begin{equation}
    h_{\rm GR}=\sum_{\ell,m}h_{\ell m}, 
    \qquad
    h_{\ell m}=\sum_{n}\sum_{p}h_{\ell mnp},
    \label{eq:mode-family-decomposition}
\end{equation}
For the SNR decomposition we introduce a compact index $i$ to label the retained mode families,
$h_i \equiv h_{\ell_i m_i}$.
The SNR matrix is then given by
\begin{equation}
    \rho_{ij}^{2}\equiv (h_i|h_j),
    \label{eq:mode-family-snr-matrix}
\end{equation}
so that
\begin{equation}
    \rho_{\rm tot}^{2}=(h_{\rm GR}|h_{\rm GR})
    =\sum_{i,j}\rho_{ij}^{2}.
    \label{eq:mode-family-total-snr}
\end{equation}
Figure~\ref{fig:mode-snr-matrix} shows the mode-family SNR for a representative eccentric LISA source with $e_0=0.4, {\cal M}=3000M_{\odot}, q=5$ observed from $t_{c}=4 {\rm yr}$ before coalescence. The plotted modes are $(\ell,m)=(2,0),(2,2),(3,0),(3,2),(4,0),(4,2),(4,4)$. The SNR budget is overwhelmingly dominated by the $\ell=2$ modes. Among them, the $(2,2)$ mode provides the largest diagonal
contribution, while the $(2,0)$ mode is also appreciable because it is
already present at Newtonian order. The combined contribution from the
remaining non-dominant modes is only $\simeq10^{-4}$ of the diagonal
SNR budget. The off-diagonal mode interference is also small: in this example the signed cross-mode contribution is about $\simeq 10^{-3}$ of $\rho_{\rm tot}^{2}$, while the sum of the absolute cross terms is about $\simeq 10^{-2}$ of $\rho_{\rm tot}^{2}$. Thus, even for an eccentric and relatively massive system, the total SNR is controlled primarily by the dominant quadrupole mode.

\begin{figure*}
\centering
\includegraphics[width=0.45\textwidth]{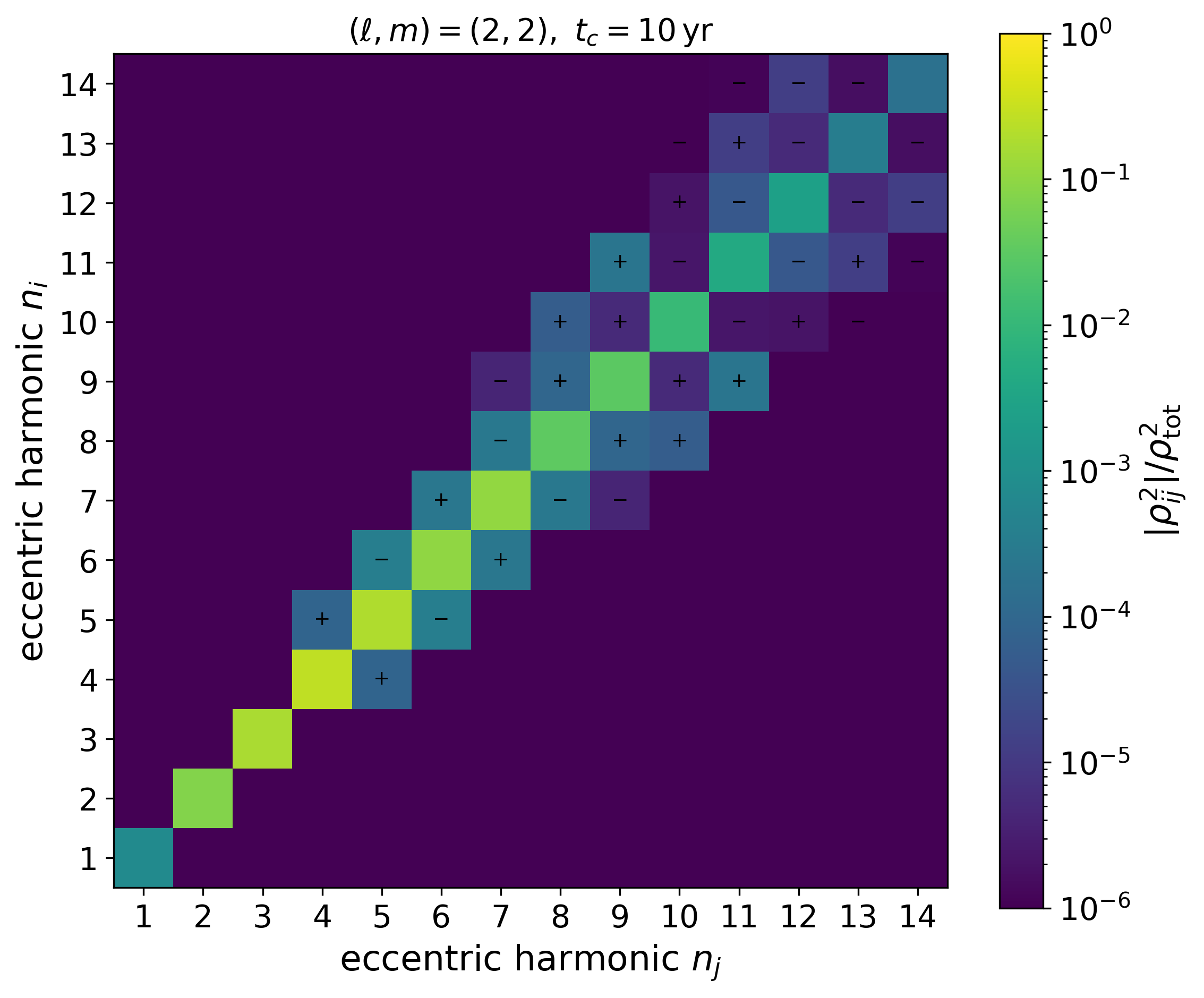}
\includegraphics[width=0.45\textwidth]{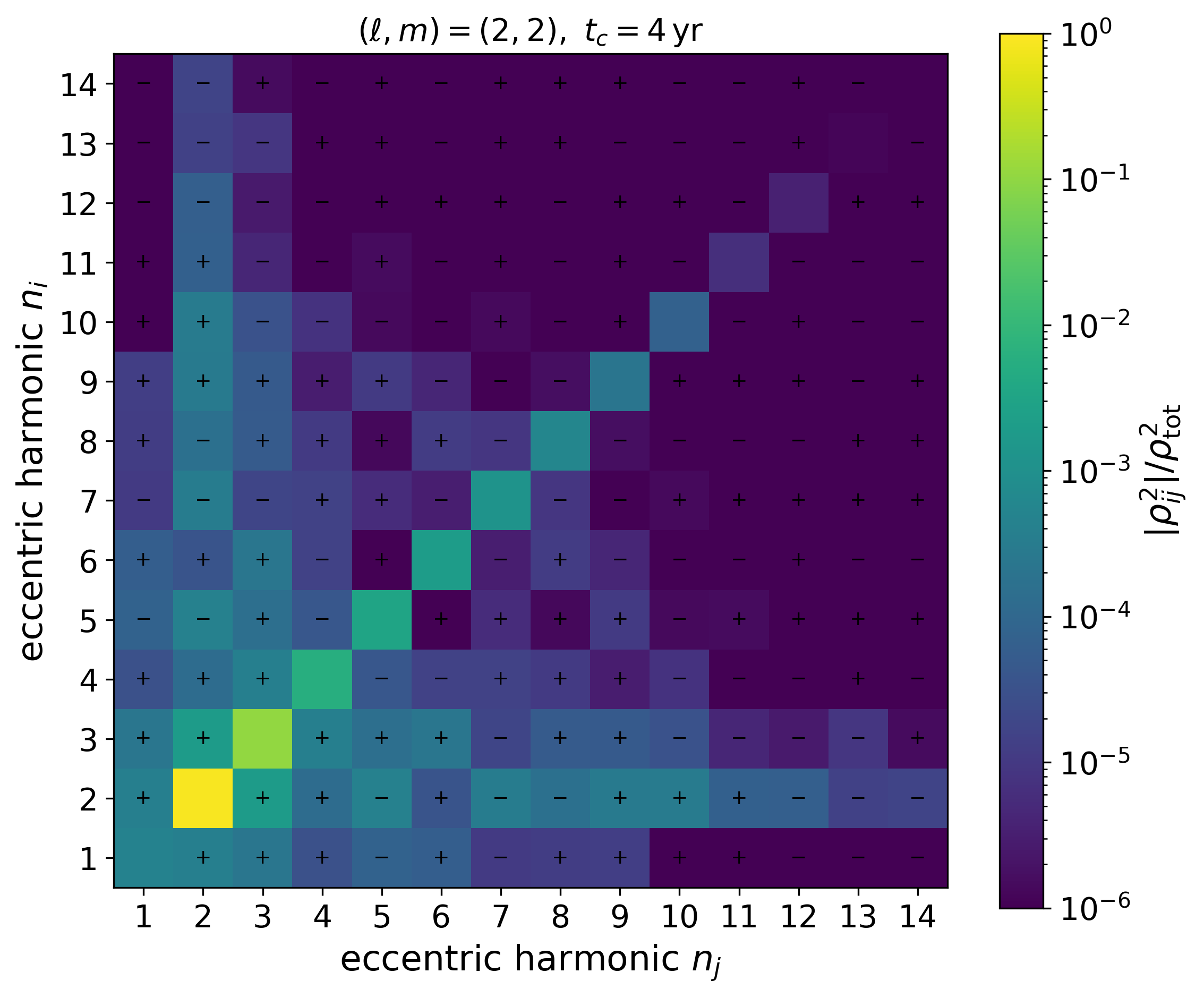}
\caption{
Fractional SNR budget of the $(\ell,m)=(2,2)$ waveform grouped by eccentric harmonic index $n$. The $+$ and $-$ symbols indicate the sign of the off-diagonal cross terms. The source parameters are $e_{0}=0.4$, ${\cal M}=3000M_{\odot}$, $q=1$, and $T_{\rm obs}=4\,{\rm yr}$. The left panel corresponds to an observation beginning $t_{c}=10\,{\rm yr}$, the right panel corresponds to $t_{c}=4\,{\rm yr}$. The diagonal harmonic contributions dominate in both cases, while cross-harmonic interference becomes more visible for the system observed closer to merger. }
\label{fig:harmonic-snr-matrix}
\end{figure*}

We next isolate the eccentric-harmonic content of the dominant quadrupole mode $(\ell,m)=(2,2)$. For this purpose, we  group the waveform by radial harmonic index,
\begin{equation}
    h_{22}=\sum_n h_n, 
    \qquad
    h_n=\sum_p h_{22np}.
    \label{eq:harmonic-decomposition}
\end{equation}
For the SNR matrix, the indices $i$ and $j$ now label eccentric harmonics, $h_i\equiv h_{n_i}$ and $h_j\equiv h_{n_j}$, and $\rho_{ij}^{2}\equiv (h_{n_i}|h_{n_j})$. The total SNR squared is
\begin{equation}
\rho_{\rm tot}^{2}
= \sum_{i}\rho^{2}_{ii}
+2\sum_{i<j}\rho^{2}_{ij}.
\label{eq:harmonic-total-snr}
\end{equation}
In the full symmetric matrix, each cross-harmonic pair appears twice; the signed pair contribution to $\rho_{\rm tot}^{2}$ is therefore $2\rho^{2}_{ij}/\rho_{\rm tot}^{2}$.

Figure~\ref{fig:harmonic-snr-matrix} compares the eccentric-harmonic SNR budget for a binary at different starting times before coalescence. We consider an equal-mass binary with $e_{0}=0.4$, ${\cal M}=3000M_{\odot}$, and $T_{\rm obs}=4\,{\rm yr}$. The left panel corresponds to an observation beginning $t_{c}=10{\rm yr}$, for which $f_{0}=5.2\times10^{-4}{\rm Hz}$. The right panel corresponds to $t_{c}=4\, {\rm yr}$, for which $f_{0}=7.3\times10^{-4}{\rm Hz}$.

The two cases show qualitatively different harmonic distributions. For the less relativistic system, the diagonal terms account for most of the full SNR-squared, with a signed cross-harmonic contribution of only $\simeq 4 \times 10^{-5}\rho_{\rm tot}^{2}$. The largest diagonal contribution comes from around $n=4$ and comparable support from neighbouring harmonics. For the $t_c=4 \,{\rm yr}$ configuration, the system is observed closer to merger and at a larger starting frequency. The harmonic content becomes more concentrated, with the $n=2$ harmonic carrying the dominant diagonal contribution and the $n=3$ harmonic providing the next largest contribution. This is partly because higher eccentric harmonics are pushed to the less sensitive regions of the LISA noise curve and spend less time in the LISA band. Some of the  cross-harmonic terms are more visible in this more relativistic case, contributing $\simeq 10^{-2}\rho_{\rm tot}^{2}$ in absolute sum. Therefore cross terms become more important when the binary is observed closer to merger.

\begin{figure*}
    \centering
    \includegraphics[width=0.95\textwidth]
    {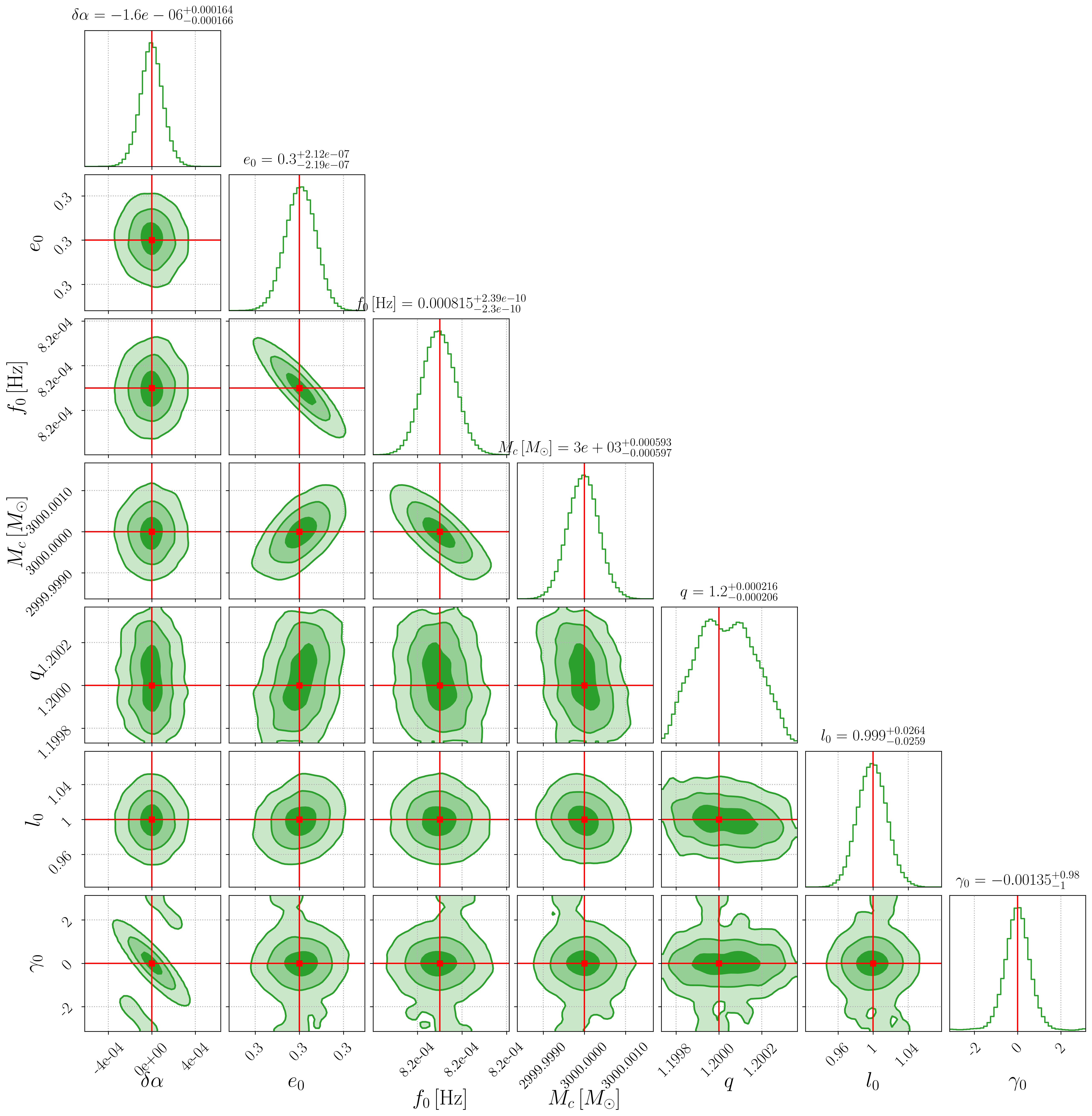}
    \caption{One- and two-dimensional marginalized posterior distributions for a GR injection recovered with the restricted waveform model. The source has detector-frame chirp mass
   $\mathcal{M}=3000\,M_\odot$, initial eccentricity $e_0=0.3$, and mass
   ratio $q=1.2$. It is initialized $t_c=4\,\mathrm{yr}$ before
coalescence and observed for $T_{\rm obs}=4\,\mathrm{yr}$, yielding a SNR of $\rho\simeq400$. Red lines indicate the
injected values.}
\label{fig:corner-sb-mass-comparison}
\end{figure*}

\section{Measurement of the non-GR parameter}
\label{sec:results}

We now assess how accurately LISA observations of eccentric binaries can constrain the phenomenological deviation parameter $\delta\alpha$. 
We use~\textsc{lisabeta} and sample the posterior distribution using
\textsc{PTemcee}~\cite{2016MNRAS.455.1919V}, a parallel-tempered Markov-chain Monte Carlo implementation built on the affine-invariant
ensemble sampler~\cite{2013PASP..125..306F}. The method evolves ensembles of walkers at a hierarchy of temperatures and allows exchanges between
adjacent temperature levels, thereby improving mixing and facilitating exploration of multimodal or strongly correlated parameter spaces. All injections considered in this section satisfy the GR prediction,
$\delta\alpha=0$, and are analyzed with the same waveform implementation used to generate them. We use the standard priors on GR parameters and uniform priors on $\delta\alpha$ (See Table~\ref{tab:priors} in Appendix~\ref{app:priors}). The restricted and full cases are treated separately. Although both
recover GR at $\delta\alpha=0$, their finite harmonic and sideband representations are not numerically identical. Their posterior widths should therefore not be interpreted as constraints obtained from two exactly identical GR waveforms.

Unless stated otherwise, the binaries have mass ratio  $q=1.2$, $t_c = 4$ yr and are observed for $T_{\rm obs}=4\,{\rm yr}$. Other extrinsic parameters are chosen randomly within their allowed ranges and are fixed for all sources. For comparisons at fixed SNR, the starting frequency is determined separately for each source so that it spans the prescribed observation interval, and the luminosity distance is adjusted to obtain the target LISA SNR. We use the same uniform prior on $\delta\alpha$ for all eccentricities within each comparison.

The eccentricity-dependent results for the restricted and full models were obtained at different target SNRs, $\rho=400$ and $\rho=50$,
respectively, so their posterior widths should not be interpreted as a direct measure of their relative sensitivity. Nevertheless, the full
model yields constraints of the same order of magnitude at an SNR lower by a factor of eight. This indicates the substantially stronger
phase response obtained when the deformation is also applied to the dominant lower-order angular carriers.

\paragraph{Restricted model.}
We first consider the restricted model, in which $\delta\alpha$ modifies only the explicit precessional sideband phases. This provides a conservative implementation because the deformation is restricted to the waveform components that explicitly resolve the secular pericenter phase.

Figure~\ref{fig:corner-sb-mass-comparison} shows the marginalized
posterior distributions for a GR injection with $e_0=0.3$,
$\mathcal{M}=3000\,M_\odot$, and SNR $\rho\simeq400$, recovered
with the restricted waveform. We show the intrinsic and phase
parameters most relevant to the measurement of $\delta\alpha$. The GR
value $\delta\alpha=0$ is recovered, and the injected values of all
shown parameters lie within their $90\%$ credible regions. The
resulting $90\%$ credible bound on the deviation parameter is
$|\delta\alpha|\lesssim1.6\times10^{-4}$.

The initial eccentricity $e_0$ and frequency $f_0$ are particularly well
constrained, with absolute posterior uncertainties of order
$10^{-7}$ and $10^{-10}\,\mathrm{Hz}$, respectively. The posterior exhibits pronounced correlations among $e_0$, $f_0$,
and $\mathcal{M}$: $f_0$ is anticorrelated with both $e_0$ and
$\mathcal{M}$, while $e_0$ and $\mathcal{M}$ are positively
correlated.

A distinct covariance is visible between $\delta\alpha$ and the initial
periastron phase $\gamma_0$. Writing the accumulated precession phase
schematically as
\begin{equation}
    \gamma_\alpha(f)= \gamma_0+ \left(1+\delta\alpha\right)
    \Delta\gamma_{\rm GR}(f),
\end{equation}
the corresponding first-order variation of a resolved sideband phase is
\begin{equation}
    \delta\Psi_{np}(f) \simeq -p\,\delta\gamma_0
    -p\,\Delta\gamma_{\rm GR}(f)\,\delta\alpha .
    \label{eq:alpha-gamma-covariance}
\end{equation}
A shift in $\gamma_0$ can therefore partially compensate for the phase
change generated by $\delta\alpha$, producing the observed
anticorrelation. The degeneracy is not exact because
$\Delta\gamma_{\rm GR}(f)$ evolves across the signal and the waveform
contains multiple harmonics and precession structures with different
values of $n$ and $p$.

\begin{figure}
    \centering
    \includegraphics[width=\columnwidth]
    {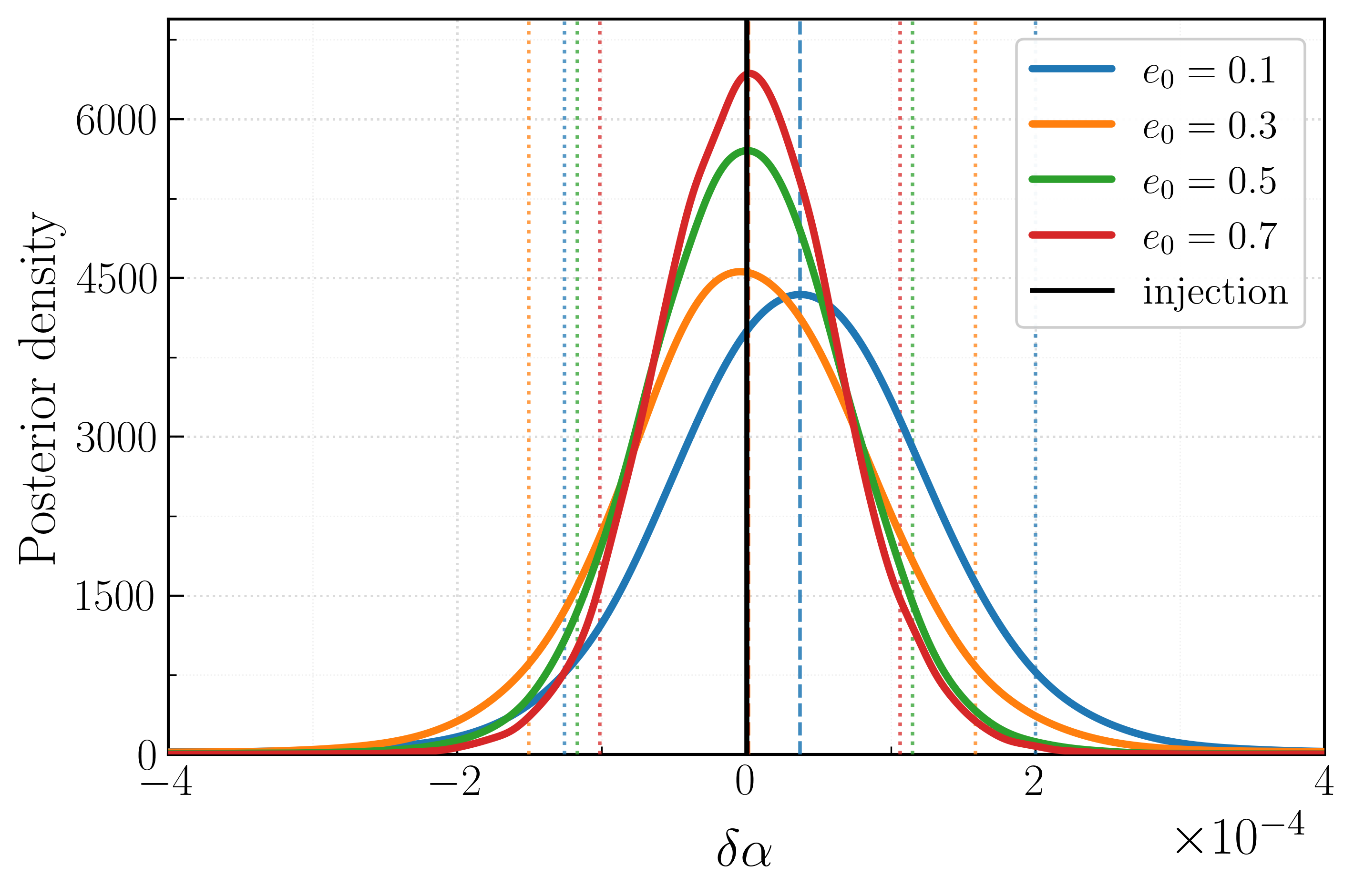}
    \caption{One-dimensional marginalized posterior distributions of $\delta\alpha$ for GR injections recovered with the restricted waveform model. The binaries have $\mathcal{M}=300\,M_\odot$, $q=1.2$,
    $T_{\rm obs}=4\,{\rm yr}$, $\rho=400$, and initial
    eccentricities $e_0=\{0.1,0.3,0.5,0.7\}$. For each eccentricity, the starting frequency is chosen consistently with the prescribed
    observation interval and the luminosity distance is adjusted to maintain the common SNR. Colored dashed lines mark posterior medians, colored dotted lines correspond to $90\%$ credible intervals, and the black vertical line marks the GR injection
    $\delta\alpha=0$. The posterior narrows with increasing eccentricity,
    showing that the additional harmonic and sideband structure improves
    the measurement of the conservative precession deformation at fixed SNR.}
    \label{fig:alpha-ecc-sb}
\end{figure}

The eccentricity dependence of the sideband measurement is shown in
Fig.~\ref{fig:alpha-ecc-sb}. We choose the lower-mass system $\mathcal{M}=300M_{\odot}$ to examine the eccentricity dependence in a regime where a large number of orbital and precession cycles are accumulated over the observation. These injections are compared at the same SNR, $\rho \approx400$. The $\delta\alpha$ posterior narrows systematically as the initial eccentricity is increased from
$e_0=0.1$ to $e_0=0.7$.

In the restricted model, the information on $\delta\alpha$ resides in the
relative phases of the explicitly precessional components. Increasing the eccentricity redistributes measurable signal power over additional radial harmonics and their associated precession sidebands. Their distinct frequency evolution supplies multiple, non-redundant measurements of the
same conservative frequency splitting. This additional structure in the waveform reduces the ability of $f_0$, $\mathcal M$, and the initial phase parameter $\gamma_0$ to mimic a change in $\delta\alpha$. This therefore leads to improved bounds on $\delta\alpha$. Quantitatively, the central $90\%$ credible interval reduces by a factor of $\approx 2$. The improvement is moderate because two competing effects are present. Increasing eccentricity adds harmonics and precession sidebands that help break parameter degeneracies,
but, in our fixed four-year comparison, the starting frequency is lowered
for the more eccentric systems. They consequently accumulate fewer secular
precession cycles during the observation, partially offsetting the gain
from the richer harmonic structure.

The broadest posterior at $e_0=0.1$ also demonstrates that high SNR alone does not eliminate intrinsic phase degeneracies at low eccentricity. High SNR supplies the overall phase precision, while eccentricity determines how effectively that precision can be separated among the conservative dynamics and the remaining source parameters.

\begin{figure*}
    \centering
    \includegraphics[width=0.95\textwidth]
    {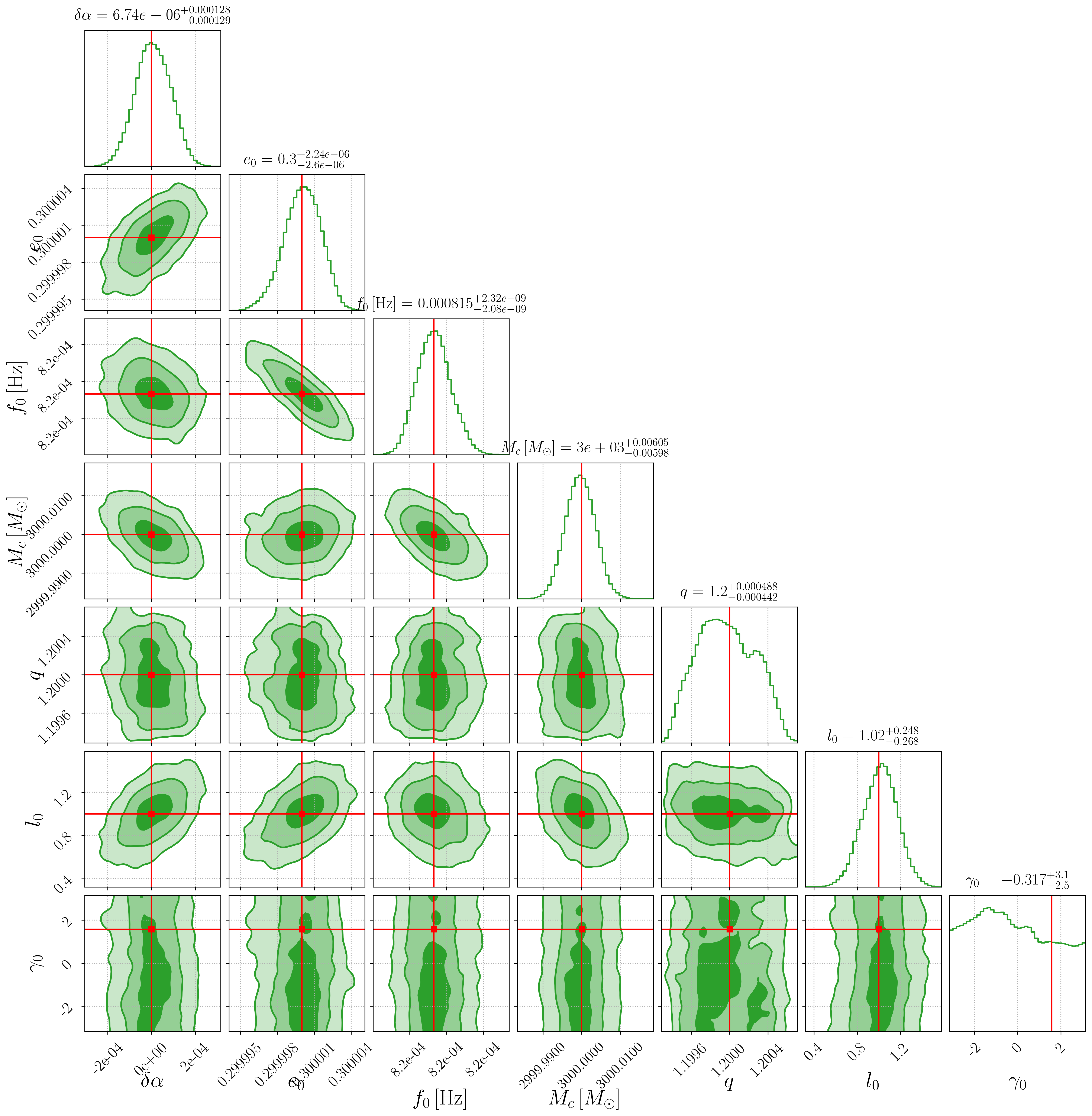}
    \caption{One and two-dimensional posterior distributions for GR
    injections recovered with the full waveform model.
    The source parameters are the same as in Fig.~\ref{fig:corner-sb-mass-comparison}
    but with the SNR $\rho=50$.}
    \label{fig:corner-cp-mass-comparison}
\end{figure*}

\paragraph{Full model.}

We next consider the full representation. Here the deformation is applied directly to the secular relation between the azimuthal and radial phases before the waveform is decomposed into a finite set of sidebands. Consequently, the dominant carrier phase itself responds to $\delta\alpha$. This is our fiducial phenomenological implementation of a deformation of the conservative frequency ratio $K=\Omega_\theta/\Omega_r$.

Figure~\ref{fig:corner-cp-mass-comparison} shows the marginalized
posterior distributions for a GR injection with
$\mathcal{M}=3000\,M_\odot$, $e_0=0.3$, and LISA SNR $\rho=50$,
recovered with the full waveform. We show only the
parameters most relevant to the measurement of $\delta\alpha$; the
full posterior distributions are provided in
Appendix~\ref{app:priors}. The injected GR value
$\delta\alpha=0$ is recovered, with a $90\%$ credible bound of
$|\delta\alpha|\lesssim 6\times10^{-4}$. In this recovery,
$\delta\alpha$ shows small covariance with $f_0$ and $q$,
whereas a positive $\delta\alpha$--$e_0$ covariance is present. The
initial periastron phase $\gamma_0$ remains only weakly constrained.

\begin{figure}
    \centering
    \includegraphics[width=0.95\columnwidth]
    {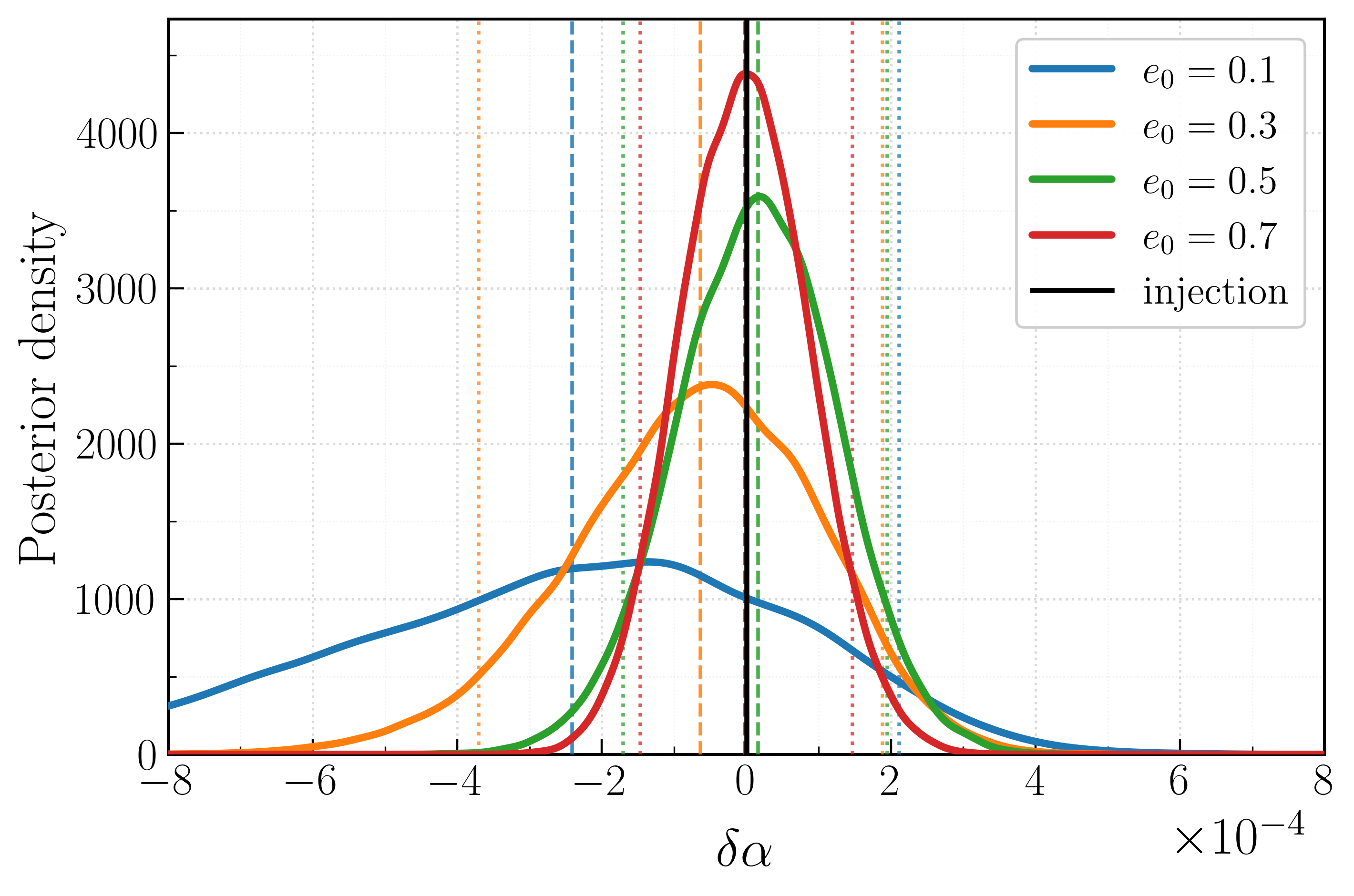}
    \caption{One-dimensional marginalized posterior distributions of $\delta\alpha$ obtained with the full waveform model for GR injections with $\mathcal M=300\,M_\odot$, SNR $\rho=50$, and initial
    eccentricities $e_0=\{0.1,0.3,0.5,0.7\}$. The posteriors become narrower with
    increasing eccentricity.
    }
    \label{fig:alpha-ecc-cp}
\end{figure}

Figure~\ref{fig:alpha-ecc-cp} shows the eccentricity dependence of the
carrier-phase measurement for $\mathcal M=300\,M_\odot$ at fixed SNR $\rho=50$. The GR value is recovered within the $90\%$
credible interval in every case. As in the sideband analysis, the
marginalized posterior becomes progressively narrower as the eccentricity increases. For $e_0=0.7$, we obtain a $90\%$ credible bound of $|\delta\alpha|\lesssim 1.4\times10^{-4}$, a factor of $\sim 5$ tighter than the corresponding bound for $e_0=0.1$.

Both waveform representations retain sensitivity to $\delta\alpha$ at
low eccentricity because the conservative frequency ratio
$K_{\alpha}=\Omega_{\theta}/\Omega_{r}$ has a smooth circular limit and
continues to enter nonvanishing waveform structures. In the
restricted model, this response is confined to the explicit
precession-dependent 1PN structures, whereas in the full model
it is also carried by the lower-order angular carriers. At fixed SNR, increasing the eccentricity produces a 
systematic improvement in the $\delta\alpha$ constraint in both
prescriptions. The additional radial harmonics and
precession-dependent structures provide nonredundant phase information
that helps disentangle $\delta\alpha$ from the orbital and
initial-phase parameters. Eccentricity therefore enhances and sharpens
the measurement. We isolate the SNR difference
under controlled conditions below using a fixed-source log-likelihood
diagnostic.

\paragraph{Likelihood-based distinguishability criterion.}
To compare the intrinsic response of the two representations
without the nuisance of different posterior volumes or target SNRs, we
compute the log-likelihood change produced by a nonzero $\delta\alpha$ for
a GR injection. With the initial pericenter phase fixed to its injected
value, we define
\begin{equation}
    \Delta\ln{\cal L}_{\rm fix}(|\delta\alpha|)=
    \ln{\cal L}(d|h_{\rm GR}) -\ln{\cal L}
    \left(d|h_{\delta\alpha,\gamma_0^{\rm inj}}
    \right),
    \label{eq:delta-lnl-fixed}
\end{equation}
where $d=h_{\rm GR}$ and all remaining source parameters are held fixed. We also construct the log-likelihood change after maximizing over the initial pericenter phase $\gamma_0$,
\begin{equation}
    \Delta\ln{\cal L}_{\rm max (\gamma_0)}(\delta\alpha)
    = \ln{\cal L}(d|h_{\rm GR}) - \max_{\gamma_0}
    \ln{\cal L}\left(d|h_{\delta\alpha,\gamma_0}\right).
    \label{eq:delta-lnl-profiled}
\end{equation}
At each fixed $\delta\alpha$, all other parameters are held fixed while
$\gamma_0$ is varied, and we retain the value that maximizes the
likelihood. This gives the smallest likelihood loss relative to the GR
injection that can be obtained by adjusting the initial pericenter
phase.

\begin{figure*}
     \centering
    \includegraphics[width= 0.95\textwidth]
    {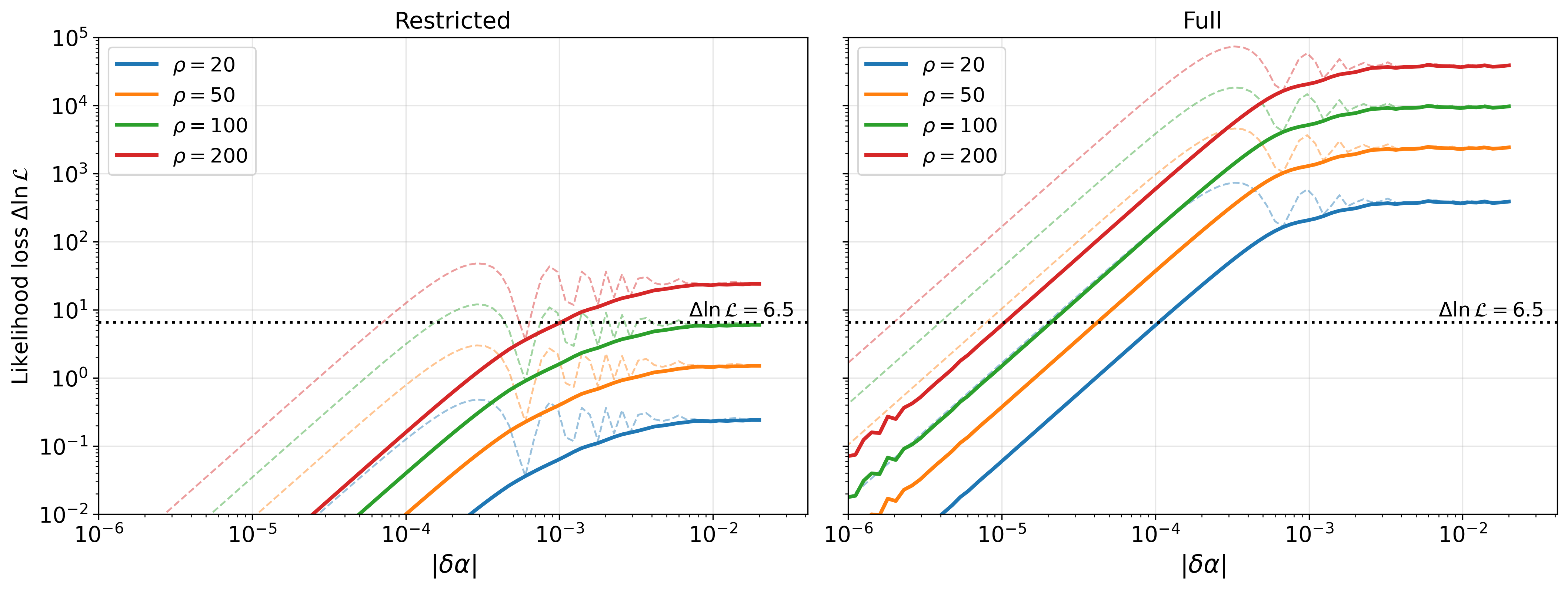}
    \caption{
Log-likelihood change induced by a nonzero $\delta\alpha$ for a GR
injection with $\mathcal{M}=3000\,M_\odot$, $q=1.2$, $e_0=0.5$,
$T_{\rm obs}=4\,{\rm yr}$, and $t_c=4\,{\rm yr}$. The left panel
shows the restricted model implementation and the right panel shows
the full waveform implementation. Curves are rescaled to the
indicated target SNRs. Solid curves show the log-likelihood change
after maximizing over $\gamma_0$ at each fixed $|\delta\alpha|$,
while dashed curves keep $\gamma_0$ at its injected value. The
horizontal dotted line marks
$\Delta\ln{\cal L}=D/2=6.5$, where $D=13$ is the dimension of the
full parameter space, and provides the corresponding
$D$-dimensional Gaussian distinguishability reference
scale. The full waveform model
reaches this threshold at substantially smaller $|\delta\alpha|$
because the dominant carrier accumulates the deformed secular phase.
}\label{fig:delta-alpha-likelihood-change}
\end{figure*}

Figure~\ref{fig:delta-alpha-likelihood-change} shows the result for a GR injection with $\mathcal M=3000\,M_\odot$, $q=1.2$, $e_0=0.5$,
$T_{\rm obs}=4\,{\rm yr}$, and $t_c=4\,{\rm yr}$. The curves are
rescaled to different target SNRs using the quadratic scaling
$\Delta\ln{\cal L}\propto\rho^2$. Following the Gaussian distinguishability criterion of Ref.~\cite{Chatziioannou:2017tdw}, we adopt
$\Delta\ln{\cal L}_{\rm th}=D/2$ as a D-dimensional Gaussian distinguishability reference scale, where
$D=13$ is the dimension of the full parameter space. This gives
$\Delta\ln{\cal L}_{\rm th}=6.5$. In the restricted model, it requires higher values of SNR to reach the threshold change in likelihood, whereas the full model respond to the $\delta\alpha$ change at much lower values of SNR. Quantitatively, for the restricted model, it requires $\vert\delta\alpha \vert \gtrsim 10^{-3}$ at $\rho = 200$ to reach the threshold change while full model requires a much smaller value $\vert\delta\alpha\vert \gtrsim 10^{-5}$ at same $\rho = 200$.  

The difference between the two implementations is substantial. In the
restricted waveform, only the explicit precession sidebands respond
to $\delta\alpha$. The affected fraction of the signal is therefore
comparatively small, and optimizing over $\gamma_0$ removes a significant
part of the phase difference. In the full waveform, the
dominant carrier phase itself is modified. The log-likelihood change
rises at much smaller $|\delta\alpha|$, and optimizing over
$\gamma_0$ cannot remove the accumulated frequency-dependent carrier-phase
difference.

At sufficiently large $|\delta\alpha|$, the curves approach plateaus.
This is because once the affected waveform components have become effectively incoherent
with the GR signal, further phase accumulation primarily produces additional
phase wrapping rather than a systematic increase in the mismatch. The
restricted model curves saturate at lower values because only a limited
fraction of the total waveform power is dephased. The full model curves
saturate at much larger values because the deformation affects a larger
fraction of the signal, including its dominant component. The oscillations
in the fixed-$\gamma_0$ curves arise from phase wrapping and interference:
different harmonics and sidebands alternately re-align and de-align as
$|\delta\alpha|$ is increased. Optimizing over $\gamma_0$ follows a smoother
lower envelope of these interference fringes.

It is important to emphasize that this figure is not a Bayesian posterior constraint. All source parameters
are fixed, apart from the maximization over $\gamma_0$ in the solid curves,
and no prior-volume effects are included. Its purpose is to isolate the
waveform-level response of the two parametrizations. The full posterior
constraints additionally depend on correlations with $e_0$, $f_0$,
$\mathcal{M}$, $q$, and the extrinsic parameters.

Taken together, the Bayesian recoveries and the controlled likelihood
diagnostic establish two complementary conclusions. The
coherent carrier or sideband phase accumulation supplies the fundamental
sensitivity to $\delta\alpha$, while eccentricity improves the measurement
by providing multiple harmonics with distinct phase evolution. High SNR
therefore supplies the overall phase precision, whereas increasing
eccentricity improves the robustness of the marginalized
constraint by breaking degeneracies. The full model is
intrinsically more responsive because the deformation acts on the dominant
secular carrier, while the restricted model provides a conservative
comparison in which the deviation is restricted to explicit precession
sidebands.

\begin{figure}
    \centering
    \includegraphics[width=0.45\textwidth]
    {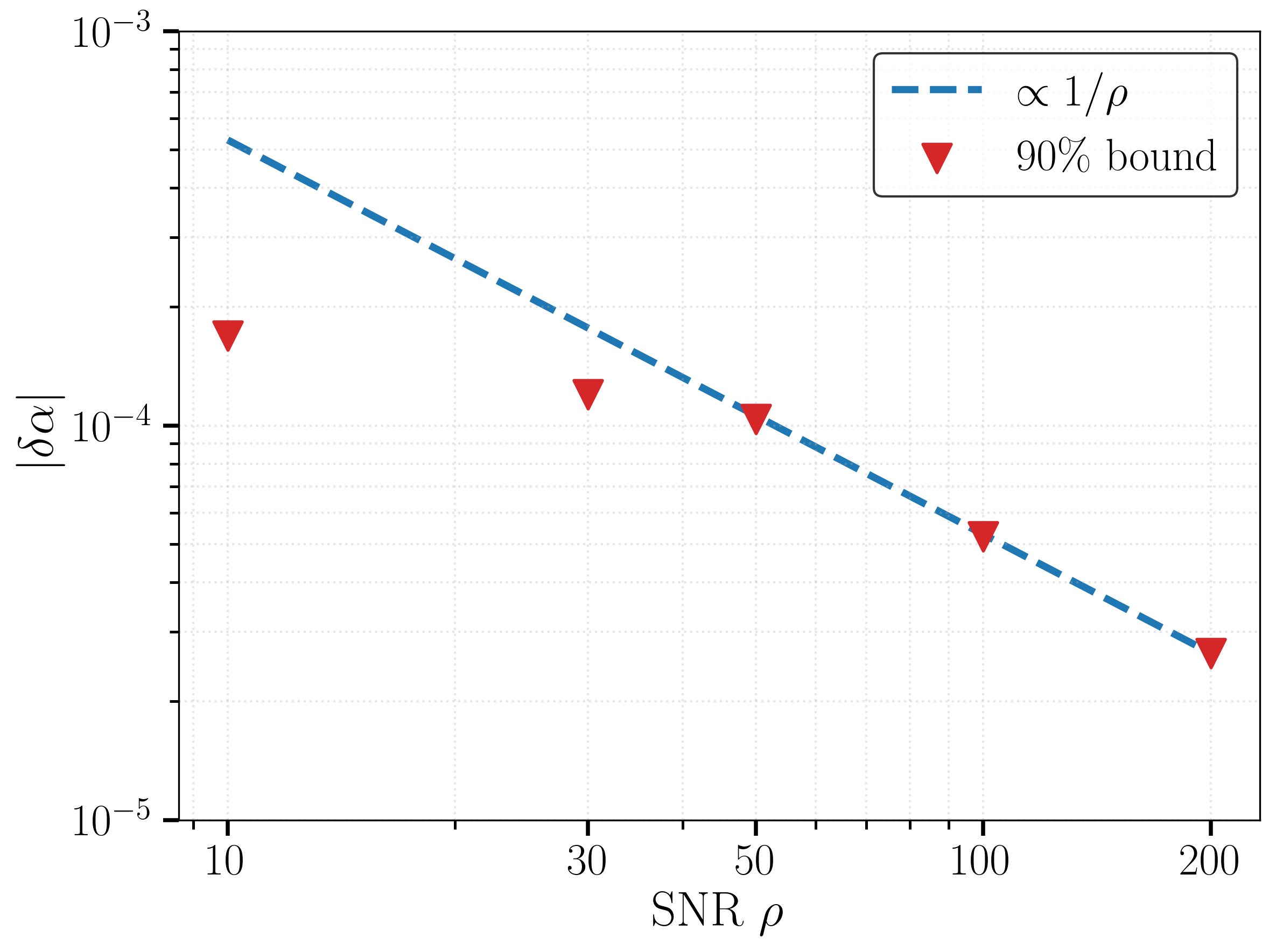}
    \caption{SNR dependence of the $\delta\alpha$ constraint for the full waveform model. The source parameters are fixed to $e_0=0.5$, $q=1.2$, and $\mathcal{M}=3000\,M_\odot$, while the injected
SNR is varied by changing $d_L$. The red triangles show
the $90\%$ credible bound, with
$\delta\alpha_{\rm GR}=0$. The dashed blue curve shows the expected asymptotic scaling $\propto 1/\rho$, normalized to the highest-SNR point. The high-SNR injections follow the $1/\rho$ scaling closely, indicating that the measurement is likelihood dominated in this regime. The lower-SNR points deviate mildly from this scaling, as expected when parameter degeneracies and non-Gaussian posterior structure become more important.
}
\label{fig:alpha-snr-scaling}
    \label{fig:B90_snr}
\end{figure}

Next, we perform full Bayesian inference for different values of SNR using the full parametrization. Figure~\ref{fig:alpha-snr-scaling} shows how the projected constraint on $\delta\alpha$ scales with the injected SNR. The intrinsic parameters are held fixed at $e_0=0.5$, $\mathcal{M}=3000\,M_\odot$, and the SNR is varied by changing the luminosity distance. We report the $90\%$ bounds, with $\delta\alpha_{\rm GR}=0$. As expected, the constraint on $\delta\alpha$ tightens systematically with increasing SNR. For $\rho=200$, we obtain a $90\%$ credible bound
of $|\delta\alpha|\lesssim 2.6\times10^{-5}$.

In the high-SNR regime, one expects the posterior to become more Gaussian and the width of a well-measured
parameter to scale approximately as $1/\rho$. The high-SNR points follow this reference scaling closely, confirming that the model behaves consistently with the expected asymptotic Fisher scaling once the accumulated phase deformation is measured with sufficient precision. At lower SNR, the bounds are somewhat weaker than a simple extrapolation from
the high-SNR regime. This departure is expected: the posterior is less
Gaussian, correlations with the intrinsic phase-evolution parameters become
more important, and the measurement is less likelihood dominated.

\begin{figure}
    \centering
    \includegraphics[width=0.45\textwidth]
    {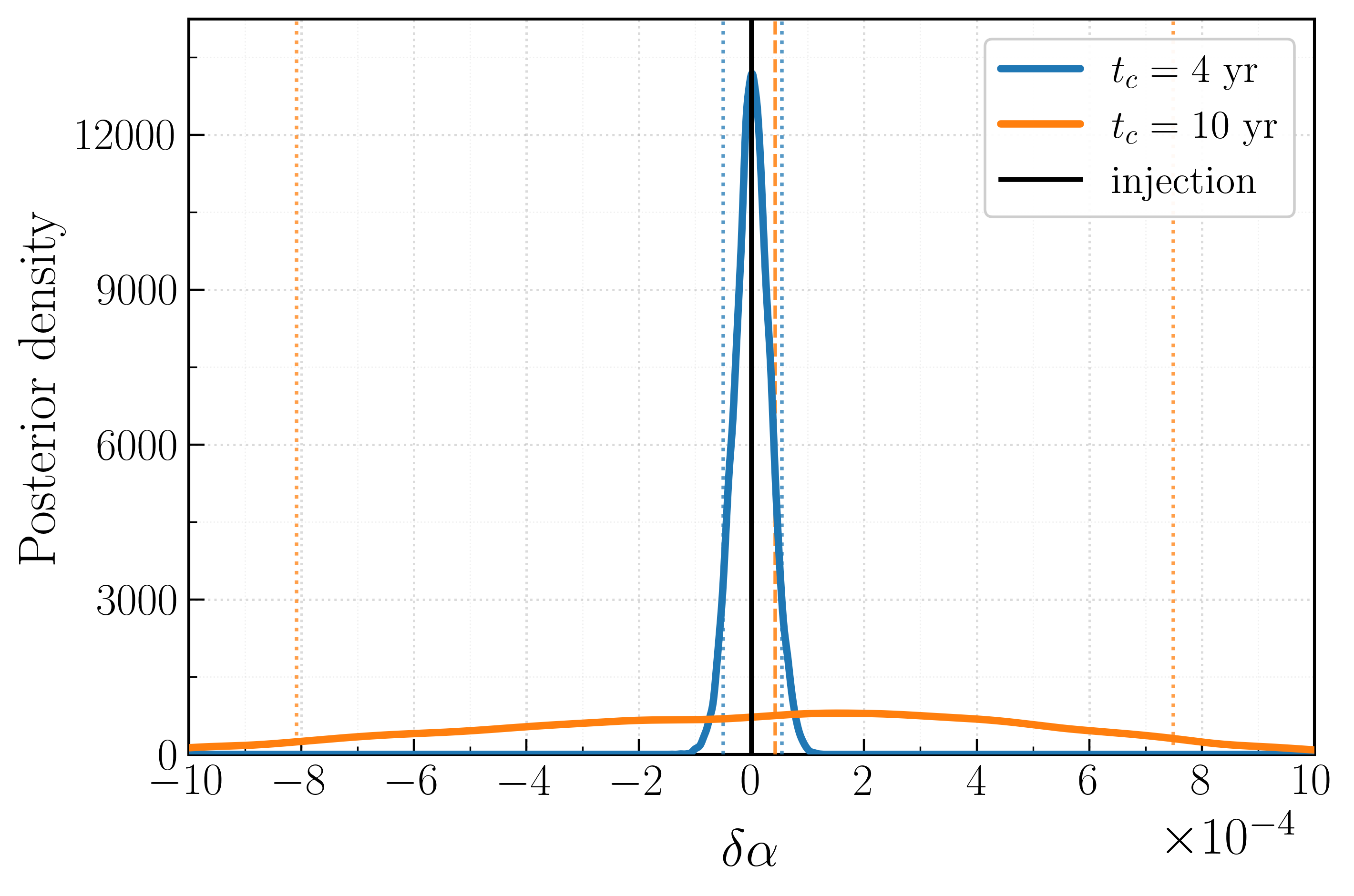}
   \caption{Marginalized posterior distributions for the parameter $\delta\alpha$ for a binary with
$\mathcal{M}=3000\,M_\odot$ and $e_0=0.5$, recovered at fixed 
SNR $\rho=100$. We compare two four-year LISA observations of the same
source parameters, beginning either $t_c=4\,{\rm yr}$ or $t_c=10\,{\rm yr}$. Although the two observations have the
same SNR, the constraint is substantially tighter for the observation
closer to coalescence. The ten-year-to-merger observation probes an
earlier, less relativistic part of the inspiral, where the
precession-induced phase structure is weaker and more degenerate with
the intrinsic binary parameters.}
\label{fig:alpha_tc_comparison}
\end{figure}

We also compare the measurement precision of $\delta\alpha$ in Figure~\ref{fig:alpha_tc_comparison} for a binary with $\mathcal{M}=3000\,M_\odot$ and $e_0=0.5$ observed at fixed SNR, but at different times before coalescence. The constraint is much tighter for $t_c = 4 $ yr binary, compared to $t_c = 10$ yr. For $t_c=4\,\mathrm{yr}$, we obtain a $90\%$ credible bound of
$|\delta\alpha|\lesssim 5\times10^{-5}$, a factor of
$\sim 15$ tighter than the corresponding bound for $t_c=10\,\mathrm{yr}$. This demonstrates that the measurability of $\delta\alpha$ is not determined by SNR alone. At earlier times the
binary is less relativistic, the periastron parameter $k_{\rm GR} \propto \nu^{2/3}$, and hence the periastron rate $\dot{\gamma}_{\rm GR} \propto \nu^{5/3}$, are smaller, so the precession-induced phase contribution accumulates
more slowly. As a result, changes in $\delta\alpha$ are more easily
absorbed by correlated shifts in the intrinsic parameters, particularly
$(e_0,f_0,\mathcal{M})$. Closer to coalescence, relativistic
precession produces a sharper secular phase signature, leading to a
substantially narrower posterior on $\delta\alpha$ at the same SNR.

It is instructive to compare our projected constraints with the
precision of binary-pulsar tests of relativistic periastron advance.
The most precise such laboratory is the Double Pulsar
PSR~J0737$-$3039A/B, for which the 16-yr timing analysis of
Ref.~\cite{PhysRevX.11.041050} measures
the change in longitude of periastron $\dot{\omega}=16.899323(13)\,\mathrm{deg\,/yr}$. The quoted
measurement uncertainty corresponds to a raw fractional precision of $\approx 10^{-6}$. At this level, however, the observed
periastron advance is no longer described by the leading 1PN
contribution alone: the 2PN correction and the Lense--Thirring
contribution associated with the spin of pulsar~A must also be
included~\cite{PhysRevX.11.041050,Hu:2020ubl}. Accounting for these
effects, the timing analysis obtains
$k_{\rm obs}/k_{\rm GR}=1.000015(26)$, corresponding to a fractional
precision of approximately $2.6\times10^{-5}$ at $1\sigma$. The
frequently quoted $1.3\times10^{-4}$ validation of GR from the same
system, reported at $95\%$ confidence, instead concerns the radiative
sector through the orbital-period decay
$\dot{P}_b$~\cite{PhysRevX.11.041050}.

The bounds forecast here,
$|\delta\alpha| \approx \mathcal{O}(10^{-4} \mbox{--} 10^{-5})$ at $90\%$ credibility for the
systems considered, are comparable to the
fractional precision of the pulsar periastron-advance test, although
the comparison is not direct because the quoted credibility levels and
parametrized quantities differ. The two measurements also probe
substantially different dynamical regimes. Our binaries evolve with a
characteristic PN velocity
$v=(M\Omega_r)^{1/3}\simeq0.03$--$0.08$ over the observation, compared
with $v\simeq2\times10^{-3}$ for the Double Pulsar, and accumulate the
corresponding phase information coherently over a multi-year
inspiral. The pulsar measurement tests the total periastron advance,
including higher-PN and spin--orbit contributions, whereas
$\delta\alpha$ parametrizes a fractional deformation of the leading
1PN conservative frequency ratio, with the radiation-reaction
evolution held fixed to its GR prediction. Spins and higher-order PN
corrections are omitted in the present analysis and will need to be
included in applications to real LISA observations. The two
measurements therefore provide complementary tests of conservative
relativistic orbital dynamics in widely separated mass and velocity
regimes.

\section{Summary and Conclusions}\label{sec:conclusion}
We have developed a parametrized null test of GR. The test introduces a dimensionless
deformation parameter $\delta\alpha$ through $K_\alpha= 1+(1+\delta\alpha)k_{\rm GR}$,
while leaving the GR radiation-reaction equations and radiative
amplitudes unchanged. We implemented this test in an eccentric frequency-domain waveform complete through 1PN order in amplitude and phase and performed Bayesian inference with the time- and frequency-dependent LISA response. This construction extends parametrized
tests of GR beyond small-eccentricity phase corrections~\cite{PhysRevD.110.124062} and makes
direct use of the multiple radial harmonics and relativistic precession
structure of eccentric signals.

A central methodological result is that the sensitivity to $\delta\alpha$ depends on how the deformation is introduced in the  waveform. In the restricted representation, the
deviation enters only the explicit precession-sideband phases. In the
full waveform representation, the same deformation is applied to
the secular radial--azimuthal phase relation before sideband
decomposition and is therefore carried by the dominant waveform
components. Both prescriptions recover GR at $\delta\alpha=0$, but they are not identical numerical GR waveforms because they organize the secular precession differently and retain different sideband sets. We present the full model results as our primary forecasts. The restricted model results are used to illustrate how strongly the constraint
changes when the same conservative deformation is restricted to the
explicit 1PN precession-sideband sector.

For the systems considered, both waveform representations yield $90\%$ bounds of $\mathcal{O}(10^{-5}-10^{-4})$, but the full waveform model
achieves comparable constraints at an SNR lower by approximately a
factor of eight. The controlled fixed-source log-likelihood comparison confirms that the stronger full model
response is intrinsic to the waveform construction: the deformation affects the accumulated phase of the loud carrier rather than only the amplitude-suppressed explicit sidebands.

The eccentricity posteriors show that SNR and eccentricity play
distinct roles. High SNR supplies the overall precision with which an
accumulated phase deformation can be resolved. Increasing eccentricity
does not merely increase the instantaneous precession rate; it
redistributes measurable power among multiple radial harmonics and
sidebands with distinct phase evolution. This additional information
breaks degeneracies between $\delta\alpha$ and other orbital and phase parameters, producing improved marginalized constraints. 

The interpretation of these bounds requires some care in the
quasicircular limit. For an exactly circular orbit, the location of
periastron and its associated angle are not independently defined.
Nevertheless, the conservative frequency ratio
$K=\Omega_{\theta}/\Omega_{r}$ has a smooth circular limit, with
$\Omega_{r}$ interpreted as the radial epicyclic frequency, and the
splitting
$\Omega_{\theta}-\Omega_{r}=k_{\rm GR}\Omega_{r}$ remains finite as
$e\rightarrow0$ \cite{Blanchet:2013haa, LeTiec:2011bk}. Since both waveform prescriptions deform this
frequency splitting, both retain a finite response to $\delta\alpha$
in the circular limit.
In the restricted prescription, this response is carried by the
surviving precession-dependent 1PN angular structures, whereas in the
full waveform prescription it is also carried by the lower-order angular carriers. The resulting bounds should therefore be
interpreted as constraints on the azimuthal--radial frequency relation
encoded in the waveform phasing, rather than as measurements of a
geometrically identifiable periastron angle. They are consequently not
in one-to-one correspondence with parametrized post-Einsteinian~\cite{Yunes:2009ke},
or related bounds for quasicircular binaries~\cite{Agathos:2013upa}, which introduce generic deformations of one or more Fourier-phase coefficients. Those coefficients generally combine conservative binding-energy and
dissipative radiation-reaction effects, whereas $\delta\alpha$ targets
the specific conservative frequency ratio
$K=\Omega_{\theta}/\Omega_{r}$ while the radiation-reaction evolution is
held fixed to its GR prediction. A quantitative comparison would
require taking the circular limit of each model, expanding the
induced phase correction in the quasicircular PN basis, and mapping it
onto the corresponding Fourier-phase coefficients.

The present analysis is intentionally controlled and has several
limitations. The waveform is complete only through 1PN order and omits
higher-PN order conservative and dissipative corrections, spins, and
orbital-plane precession. We use zero-noise injections and recover each signal with the same waveform model used to generate it. Moreover,
$\delta\alpha$ is a phenomenological deformation and has not been mapped
to the coupling constants of a specific alternative theory of gravity~\cite{Toubiana:2020vtf}. These limitations mean that the numerical bounds should be interpreted as
forecasts within a controlled waveform model, rather than as final LISA
sensitivities.

The framework is nevertheless readily extensible and can be implemented directly in more complete phenom precessing-eccentric inspiral models such as pyEFPE~\cite{Morras:2025nlp}, models containing higher PN corrections, higher modes~\cite{Morras:2026fho}, effective-one-body (EOB) waveforms~\cite{Gamboa:2024imd, Gamboa:2024hli, Gamboa:2026jht}, and hybrid eccentric waveforms~\cite{Paul:2024ujx}. Such implementations will permit injection--recovery studies across different waveform families and a quantitative assessment of higher-PN and finite-truncation systematics. Further extensions of this work include spinning binaries, environmental perturbations, multiband
LISA--ground-based observations~\cite{Barausse:2016eii,Gnocchi:2019jzp, Carson:2019rda, Gupta:2020lxa, Klein:2022rbf}, and mappings of the measured frequency-ratio deformation onto specific theories of gravity~\cite{Tahura:2018zuq}.

The framework developed here is not limited to a particular black-hole mass scale or astrophysical formation channel. In principle, the same
parametrized test can be applied to eccentric black-hole binaries across a broad mass range, from stellar-origin systems to massive black-hole
binaries~\cite{Buscicchio:2021dph,Nishizawa:2016jji,Klein:2022rbf,Garg:2023lfg,Wang:2023tle,Saini:2023wdk}, provided that the observed portion of the signal remains in the weak-field, inspiral-dominated regime. Eccentric systems are particularly well suited to this test because their multiple orbital harmonics and
relativistic precession sidebands encode the radial and azimuthal orbital
dynamics through several distinct phase contributions.

Beyond the numerical forecasts, the main conclusion
is general: a phenomenological deviation parameter placement within the waveform representation also determines the information that an observation can extract. Eccentric LISA binaries provide both the long phase baseline and
the multi-harmonic structure needed to expose this distinction. They therefore offer a powerful laboratory for precision tests of gravity.

\section{acknowledgements}
We are grateful to Martin Pessah for useful discussions. P.S. and J.S. are supported by the Villum Fonden grant No. 29466, and by the ERC Starting Grant no. 101043143 – BlackHoleMergs. P.S. and S.M. acknowledge support from the Astrophysics Center for Multi-messenger Studies in Europe (ACME), funded under the European Union’s Horizon Europe Research and Innovation Program, Grant Agreement No. 101131928. J.T. acknowledges support from the Alexander von Humboldt Foundation under the project no. 1240213 - HFST-P. L.Z. is supported by the European Union’s Horizon 2024 research and innovation program under the Marie Sklodowska-Curie grant agreement No. 101208914. The Center of Gravity is a Center of Excellence funded by the Danish National Research Foundation under grant No. 184. The Tycho supercomputer hosted at the SCIENCE HPC center at the University of Copenhagen was used for computation in this study. 

This study made use of the following software packages: {\tt lisabeta}~\cite{PhysRevD.103.083011}, {\tt ptemcee}~\cite{2016MNRAS.455.1919V}, 
{\tt Scipy}~\citep{2020NatMe..17..261V},
{\tt NumPy}~\citep{2020Natur.585..357H}, {\tt Matplotlib}~\citep{2007CSE.....9...90H}, {\tt jupyter}~\citep{soton403913}, {\tt pandas}~\citep{mckinney-proc-scipy-2010}, {\tt corner}~\cite{corner}.

\section*{DATA AVAILABILITY}
The data supporting the findings of this article will be made available upon reasonable request to the corresponding author.

\appendix
\section{Relation to the generalized-Hansen representation}
\label{app:hansen-mapping}
Here we clarify the relation between the resolved $(n,p)$
representation used in the main text and the generalized-Hansen
representation of Ref.~\cite{1PN_waveform}. We use $l$ for the mean
anomaly and reserve $(\ell,m)$ for the spherical-harmonic indices and
$\mathcal{M}$ for the chirp mass. To avoid confusion with the
periastron-advance parameter $k_{\rm GR}$, we denote by $j$ the integer
mean-anomaly harmonic denoted by $k$ in
Ref.~\cite{1PN_waveform}. We also use $s$ for the integer
true-anomaly harmonic denoted by $m$ in that reference.

Reference~\cite{1PN_waveform} introduces the drift anomaly
\begin{equation}
    \varphi' = K_{\rm GR}\varphi = \left(1+k_{\rm GR}\right)\varphi,
    \label{eq:app-drift-anomaly}
\end{equation}
where $\varphi$ is the generalized true anomaly. Starting from the
1PN azimuthal relation in Eq.~\eqref{eq:theta-motion}, the angular
dependence on $\theta$ is then reorganized in terms of $\varphi$ and
the secularly drifted anomaly $\varphi'$, with the remaining periodic
1PN correction treated separately. The true-anomaly
harmonic and the drift-anomaly contribution can be combined into a
single, generally noninteger angular order
$\lambda_{sp}^{\rm GR}$, defined by
\begin{equation}
    s\varphi+p\varphi' = \lambda_{sp}^{\rm GR}\varphi,
    \qquad  \lambda_{sp}^{\rm GR} = s+pK_{\rm GR},
    \label{eq:app-angular-order}
\end{equation}
where $p\in\{0,\pm2,\pm4\}$ labels the corresponding drift-anomaly
structure. Thus, the quantities denoted by $m_{2\pm}$ and $m_{4\pm}$
in Ref.~\cite{1PN_waveform} correspond to
\begin{equation}
    m_{2\pm} = s\pm2K_{\rm GR},
    \qquad m_{4\pm} = s\pm4K_{\rm GR}.
    \label{eq:app-mfv-orders}
\end{equation}
Because $K_{\rm GR}$ is generally noninteger, the generalized angular
orders $\lambda_{sp}^{\rm GR}$, including $m_{2\pm}$ and $m_{4\pm}$,
are also generally noninteger. These quantities enter as the angular
orders of the generalized Hansen coefficients.

The corresponding true-anomaly factors are expanded in integer
harmonics of the mean anomaly according to
\begin{align}\label{eq:app-hansen-expansion-1}
    e^{i\lambda\varphi} =  \sum_{j=-\infty}^{\infty}
    X_j^{0,\lambda}(e_r,e_t)e^{ijl}, \\
    X_j^{0,\lambda}  = \frac{1}{2\pi} \int_{-\pi}^{\pi}
    e^{i\lambda\varphi(l)-ijl}\,dl.
    \label{eq:app-hansen-expansion-2}
\end{align}
This explains why the final waveform in Eq.~(38) of
Ref.~\cite{1PN_waveform} contains only integer mean-anomaly harmonics
$jl$. The periastron-precession dependence enters through the
generally noninteger order $\lambda$ of the generalized Hansen
coefficients $X_j^{0,\lambda}$ and hence through the complete
coefficients $\mathcal{C}_j$ and $\mathcal{S}_j$ multiplying the
corresponding cosine and sine harmonics.

Our parametrized deformation is introduced through
\begin{align}\label{eq:app-deformed-order-1}
    K_{\rm GR} \rightarrow
    K_{\alpha} &=
    1+\left(1+\delta\alpha\right)k_{\rm GR}, \\
    \lambda_{sp}^{\rm GR}
    \rightarrow \lambda_{sp}^{\alpha} &= s+pK_{\alpha}.
    \label{eq:app-deformed-order-2}
\end{align}
No independent deformation is applied to the GR radiative prefactors.
Nevertheless, the resulting complex harmonic coefficients depend on
$\delta\alpha$, because the generalized Hansen coefficients depend on
the deformed angular order $\lambda_{sp}^{\alpha}$. In the numerical
waveform, Eqs.~\eqref{eq:app-hansen-expansion-1}-\eqref{eq:app-hansen-expansion-2} are evaluated directly by
discretizing its defining integral over the mean anomaly.

The connection with the notation used in the main text follows by
writing, for fixed orbital elements,
\begin{equation}
    s\varphi+p\varphi'_{\alpha} = (s+p)\varphi+p\Delta\gamma_{\alpha},
    \label{eq:app-resolved-decomposition}
\end{equation}
where $\Delta\gamma_{\alpha} = \varphi'_{\alpha}-\varphi$ and $\varphi'_{\alpha}  =  K_{\alpha}\varphi.$ The first term in Eq.~\eqref{eq:app-resolved-decomposition} is periodic
over the radial motion and can be expanded in integer mean-anomaly
harmonics, whereas the second identifies the secular precession
structure. During the adiabatic inspiral, $\gamma_{\alpha}$ is
understood as the accumulated phase satisfying
$\dot{\gamma}_{\alpha}=k_{\alpha}\dot{l}$. This motivates the resolved
phase $nl+p\gamma_{\alpha}$ used in the main text.

The two waveform prescriptions differ in the angular structures to
which Eqs.~\eqref{eq:app-deformed-order-1}-\eqref{eq:app-deformed-order-2} are applied. In
$\mathcal{H}_{\rm R}$, the deformation is applied only to the explicit
1PN precession-dependent structures, with
$p=0,\pm2,\pm4$. In $\mathcal{H}_{\rm F}$, the same secular relation
is additionally assigned to the lower-order angular carriers,
producing $p=\pm2$ in the Newtonian-amplitude and
$p=\pm1,\pm3$ in the half-PN-amplitude.

Finally, the complex-exponential notation used in this work is
equivalent to the sine--cosine representation of
Ref.~\cite{1PN_waveform}, since
\begin{equation}
    \mathcal{C}_j\cos(jl)+\mathcal{S}_j\sin(jl) =
    \frac{\mathcal{C}_j-i\mathcal{S}_j}{2}e^{ijl}
    +  \frac{\mathcal{C}_j+i\mathcal{S}_j}{2}e^{-ijl}.
    \label{eq:app-complex-basis}
\end{equation}
This is only a change of harmonic basis. The explicit $(n,p)$ notation
exposes the physical radial and precessional content, whereas the
generalized-Hansen representation absorbs the drift dependence into
the complex coefficients. Consequently, an individual integer
mean-anomaly harmonic $j$ in the generalized-Hansen representation
need not map one-to-one onto an individual resolved pair $(n,p)$.

\paragraph{Numerical treatment of the orbital evolution.}
A further difference from the analytic implementation of
Ref.~\cite{1PN_waveform} concerns the construction of the orbital
evolution. In obtaining closed analytic expressions for the time and
phase functions, that reference approximates the ordinary
hypergeometric function entering $b_{\rm PN}(e)$ by unity. We do not
make this approximation. Instead, we evaluate
\begin{equation}
    {}_2F_1\!\left(  \frac{870}{2299}, \frac{13}{19};
    \frac{32}{19}; -\frac{121}{304}e^2 \right)
    \label{eq:app-hypergeometric}
\end{equation}
directly and retain its full eccentricity dependence in the 1PN
frequency--eccentricity relation $\nu(e)$ introduced in
Eq.~\eqref{eq:nu-of-e-1pn}. The time and mean-anomaly evolution are
then constructed by numerical quadrature,
\begin{equation}
    t(e)-t_0 =
    \int_{e_0}^{e}\frac{de'}{\dot e(e')},
    \qquad l(e)-l_0  =\int_{e_0}^{e} \frac{2\pi\nu(e')}{\dot e(e')}\,de'.
    \label{eq:app-numerical-evolution}
\end{equation}
The resulting functions are tabulated on an eccentricity grid and
interpolated during waveform generation. This numerical treatment
retains the full eccentricity dependence of the adopted 1PN orbital
evolution while leaving unchanged the definition of the parametrized
conservative deformation.

\begin{figure}
    \centering
    \includegraphics[width=0.95\columnwidth]
    {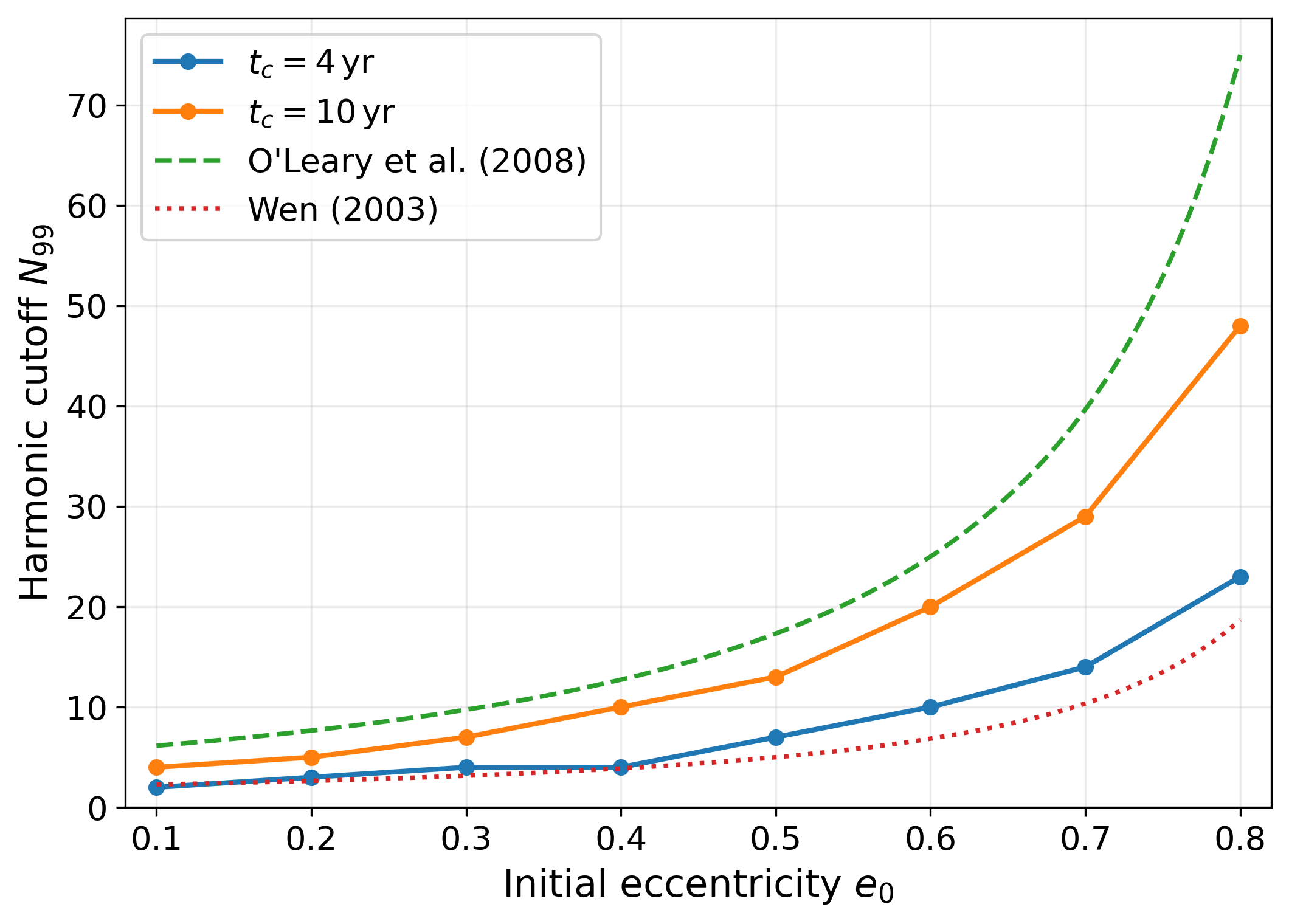}
   \caption{Minimum radial-harmonic cutoff $N_{99}$ required to retain $99\%$ of the cumulative detector-weighted $(\ell,m)=(2,2)$ diagonal SNR-squared
for an equal-mass binary with $\mathcal{M}=3000\,M_\odot$. The solid curves show the numerically computed $N_{99}$. The dashed curve shows the conservative
O'Leary et al.~\cite{OLeary:2008myb} prescription for an approximate $99\%$-power cutoff,
while the dotted curve shows the Wen~\cite{Wen:2002km} estimate of the peak-power
harmonic. The required cutoff increases with eccentricity and is
systematically larger for the observation beginning farther from
merger. The Wen estimate underestimates the detector-weighted cutoff for the highly eccentric systems considered here, whereas the O'Leary prescription remains conservative.}
\label{fig:n99-eccentricity}
    \label{fig:corner-full-sb-mc3000-e03}
\end{figure}

\section{Convergence of the radial-harmonic expansion}
\label{app:harmonic-convergence}
To determine a radial-harmonic cutoff $n_{\rm max}$ sufficient to capture the relevant signal power, we test the convergence of the radial-harmonic expansion. We compare the two prescriptions given in Eqs.~\eqref{eq:nmax-oleary} and ~\eqref{eq:npeak-wen} shown in the main text. We consider an equal-mass binary with
$\mathcal{M}=3000\,M_\odot$ and compare two four-year LISA
observations: one with $t_c = 4$ yr and one with $t_c = 10 $ yr. The convergence test is restricted to the dominant spherical-harmonic mode $(\ell,m)=(2,2)$. For each radial harmonic $n$, all retained precession sidebands are summed coherently,
\begin{equation}
    h_n = \sum_{p}h_{22np},
\end{equation}
and we define its detector-weighted self-SNR contribution as
\begin{equation}
    \rho_n^2 =  (h_n|h_n)  = \sum_{p,p'}
    (h_{22np}|h_{22np'}).
\end{equation}
Here we omit the cross terms between distinct radial harmonics so that the cumulative quantity remains monotonic. The cumulative diagonal SNR-squared fraction is
\begin{equation}
    F_N=
    \displaystyle\sum_{n=1}^{N}\rho_n^2 \bigg{/}
    {\displaystyle\sum_{n=1}^{N_{\rm ref}}\rho_n^2},
    \label{eq:cumulative-harmonic-snr}
\end{equation}
where we take $N_{\rm ref}=120$ for the reference harmonic cutoff. We then
define $N_{99} = \min\left\{ N\,:\,F_N\geq0.99 \right\}$.
The resulting $N_{99}$ is therefore a detector-weighted
harmonic-content diagnostic. It is not identical to a $99\%$ emitted
power criterion, nor does it by itself establish the waveform accuracy
required for parameter estimation at a specified SNR.

Figure~\ref{fig:n99-eccentricity} shows that the required harmonic
cutoff increases rapidly with the initial eccentricity. It also depends
strongly on where the finite observation begins relative to merger. At
fixed $e_0$, the four-year observation beginning at $t_c = 10$ yr requires substantially more harmonics than the observation
covering the final four years with $t_c = 4$ yr. The earlier system begins at a lower
orbital frequency and undergoes less circularization during the
observed interval. Consequently, a larger set of high-order radial
harmonics contributes appreciably within the LISA band. For example, at $e_0=0.8$ we find
$N_{99}=23\quad\text{for}\quad t_c=4\, {\rm yr}$ and
$N_{99}=48 \quad\text{for}\quad t_c=10\, {\rm yr}$. 
The Wen prescription~\cite{Wen:2002km} gives
$n_{\rm peak}^{\rm Wen}\simeq18$ at the same eccentricity and therefore
substantially underestimates the cutoff required by the present
detector-weighted criterion, especially for the earlier observation.
This does not indicate a failure of the Wen fit: it estimates the
peak-power harmonic rather than the upper harmonic needed to retain
$99\%$ of the cumulative signal.

The O'Leary prescription~\cite{OLeary:2008myb} lies above the numerical $N_{99}$ values over the full eccentricity range considered and therefore provides a
conservative cutoff. Although it retains more harmonics than are
strictly required by the $99\%$ diagonal-SNR criterion, this choice
avoids prematurely discarding appreciable eccentric-harmonic content.
We consequently use O'Leary Eq.~\eqref{eq:nmax-oleary} in the production
waveforms.

The $N_{99}$ criterion should nevertheless be interpreted as a
waveform-content check rather than a complete parameter-estimation
accuracy requirement. For sufficiently loud sources, convergence may
require retaining substantially more than $99\%$ of the signal norm.
A direct mismatch or log-likelihood convergence calculation would be
needed to quantify possible parameter biases associated with a
particular harmonic truncation.

\section{Priors and full corner plot}
\label{app:priors}
\begin{figure*}
    \centering
    \includegraphics[width=0.96\textwidth]
    {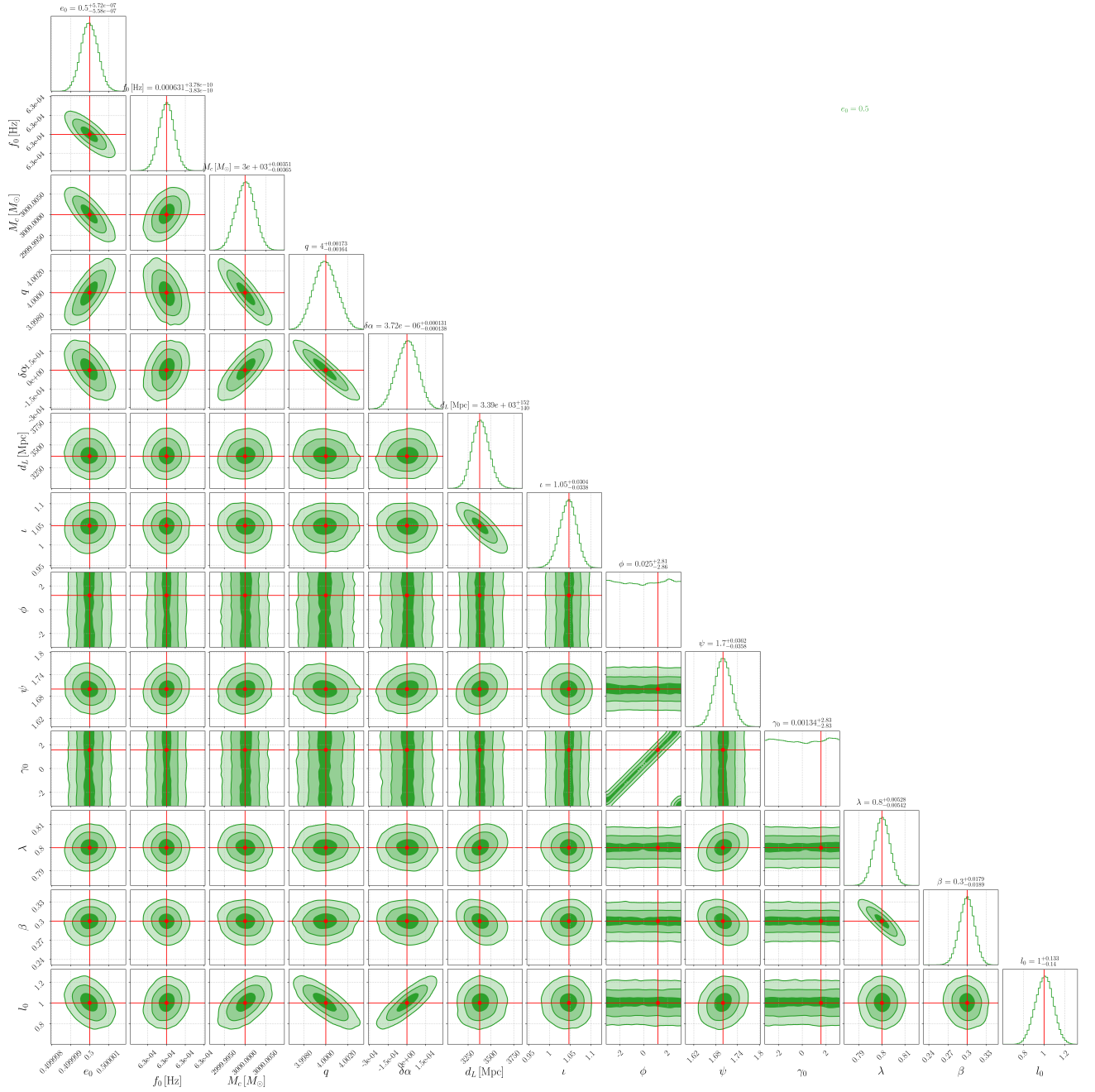}
    \caption{
    Full 13-dimensional corner plot for a GR injection recovered with
    the full waveform model. The source has
    $\mathcal{M}=3000\,M_\odot$, $e_0=0.5$, $q=4$, is observed for
    $T_{\rm obs}=4\,{\rm yr}$, and has SNR $\rho=100$.
     Red lines mark the injected values, and the green
    shaded regions show the joint posterior credible contours. Correlations between $\delta \alpha$ and most parameters are weak, with a more pronounced correlation with $q$, while the extrinsic parameters show the expected
    $\iota-d_L$ and $\gamma_0 - \phi$ correlations.
    }
    \label{fig:corner_full_carrier}
\end{figure*}

For completeness, we show the full 13-dimensional corner plot in Fig.~\ref{fig:corner_full_carrier} for the
$\mathcal{M}=3000\,M_\odot$, $e_0=0.5$ and $q=4$ system with $t_c = 4$ yr using the full model.
The sampled parameters are $\left\{e_0,\, f_0,\, \mathcal{M},\, q,\, \delta\alpha,\, d_L,\, \iota,\, \phi,\, \psi,\, \gamma_0,\, \lambda,\,  \beta,\,  l_0 \right\}.$ The priors used on waveform parameters are shown in Table~\ref{tab:priors}. Since the intrinsic parameters are measured with exquisite precision in LISA, we use wide enough priors so that it does not affect the posterior distributions. The full recovery confirms that the injected GR value
$\delta\alpha=0$ lies within the posterior support. It also shows that the
parameters most directly relevant to the $\delta\alpha$ measurement are
already captured by the reduced corner plot in the main text.

For this $q=4$ injection, the correlation between $\delta\alpha$ and $\gamma_0$ is weak while a stronger correlation with $q$ is visible. The remaining extrinsic parameters are only weakly correlated with $\delta\alpha$. The full posterior also displays the standard $\iota - d_L$ and $\phi-\gamma_0$
correlations. 

\begin{table}
\centering
\renewcommand{\arraystretch}{1.2}
\caption{Priors used for parameter estimation. $\mathcal{U}(\rm{min}, \rm{max})$} refers to the uniform distribution.
\begin{tabular}{cc}
\toprule
Parameter & Prior \\
\midrule
Chirp mass $\mathcal{M}\,(M_{\odot})$ 
    & Uniform \\
Mass ratio $q$ 
    & Uniform \\
Initial eccentricity $e_0$ 
    & Uniform \\
Initial GW frequency $f_0=f_{22}^{\rm start}$ 
    & Uniform \\
Initial pericenter angle $\gamma_0\,(\mathrm{rad})$ 
    & $\mathcal{U}(-\pi,\pi)$ \\
Non-GR parameter $\delta\alpha$ 
    & Uniform \\
Luminosity distance $d_L\,(\mathrm{Mpc})$ 
    & Uniform \\
Inclination $\iota\,(\mathrm{rad})$ 
    & $\cos\iota \sim \mathcal{U}(-1,1)$ \\
Source longitude $\lambda\,(\mathrm{rad})$ 
    & $\mathcal{U}(-\pi,\pi)$ \\
Source latitude $\beta\,(\mathrm{rad})$ 
    & $\sin\beta \sim \mathcal{U}(-1,1)$ \\
Polarization angle $\psi\,(\mathrm{rad})$ 
    & $\mathcal{U}(0,\pi)$ \\
Observer's azimuthal phase $\phi\,(\mathrm{rad})$ 
    & $\mathcal{U}(-\pi,\pi)$ \\
Initial mean anomaly $l_0\,(\mathrm{rad})$ 
    & $\mathcal{U}(-\pi,\pi)$ \\
\bottomrule
\end{tabular}
\label{tab:priors}
\end{table}

\bibliography{ref-list}
\end{document}